\documentclass{aa}
\usepackage{graphics}
\usepackage{epsfig,times}

\makeatletter

\def\sax{{\it BeppoSAX }}

\def\chiq{$\chi^2$}
\def\bb{{\sc bb}}
\def\dbb{{\sc dbb}}

\def\usc{{\sc usc}}
\def\pl{{\sc pl}}
\def\vphabs{{\sc vphabs}}
\def\wabs{{\sc wabs}}
\def\compst{{\sc compst}}
\def\gaussian{{\sc gaussian}}   
\def\comptt{{\sc comptt}}
\def\compbb{{\sc compbb}}
\def\ktin{kT_{\rm in}}

\def\ktbb{kT_{\rm bb}}
\def\kte{kT_{\rm e}}

\def\rin{R_{\rm in}}
\def\reff{kT_{\rm eff}}

\def\xspec{{\sc xspec}}
\def\s1{{s$^{-1}$ }}

\def\@cite#1#2{(#1\if@tempswa , #2\fi)}
\def\@citex[#1]#2{\if@filesw\immediate\write\@auxout{\string\citation{#2}}\fi
  \def\@citea{}\@cite{\@for\@citeb:=#2\do
    {\@citea\def\@citea{;\penalty\@m\ }\@ifundefined
       {b@\@citeb}{{\bf ?}\@warning
       {Citation `\@citeb' on page \thepage \space undefined}}%
\hbox{\csname b@\@citeb\endcsname}}}{#1}}
\makeatother  

\begin{document}

\title{{\it BeppoSAX} observations of two unclassified LMXBs: 
X1543--624 and X1556--605}

\author{R. Farinelli\inst{1}, F. Frontera\inst{1,2}, N. Masetti\inst{2}, 
L. Amati\inst{2}, C. Guidorzi\inst{1}, M. Orlandini\inst{2}, 
E. Palazzi\inst{2}, A.N. Parmar\inst{3}, L. Stella\inst{4}, M. Van der Klis \inst{5} and S.N. Zhang\inst{6}
}

\institute{$^1$Dipartimento di Fisica, Universit\'a di Ferrara, via Paradiso
12, I-44100 Ferrara, Italy\\ 
$^2$Istituto Astrofisica Spaziale e Fisica Cosmica, Sezione di Bologna, CNR,
Via Gobetti 101, I-40129 Bologna, Italy\\
$^3$Astrophysics Division, Space Science Department of ESA, ESTEC, P.O. Box
299, 2200 AG Noordwijk, The Netherlands \\
$^4$Osservatorio Astronomico di Roma, Via Frascati, 33, 00040 Monteporzio 
Catone, Italy \\
$^5$Astronomical Institute ``Anton Pannekoek'', University of Amsterdam and
Center of High Energy Astrophysics, Kruislaan 403, NL 1098 SJ Amsterdam, The
Netherlands \\
$^6$Department of Physics, University of Alabama in Huntsville, Huntsville, 
AL 35899, USA
}

\date{Received 6 May 2002/Accepted 8 January 2003}

%\thesaurus{6(02.01.2; 08.09.2; 08.14.1; 13.25.3; 13.25.5)}

\abstract{
 Observations of two unclassified Low Mass X--ray Binaries,
X1543--624 and X1556--605, are presented.
In the 2--10 keV band the first of the two sources is a factor of two
stronger than the other.
Both sources do not show X--ray bursts, dips or eclipses in their X--ray 
light curves.
We find that both spectra  are described by a two--component model 
consisting of emission from a cool accretion disk plus a Comptonized blackbody 
with $\ktbb \sim 1.5$ keV
in a low opacity plasma. The spectrum of X1543--624 
hardens from the first to the second observation, when the source slowly 
moves from right to left in the colour--colour diagram. The spectrum of 
X1556--605 can also be described by a model consisting of a blackbody plus 
 an unsaturated Comptonization with electron energy $kT_e \sim 4$ keV.
In the first observation, X1543--624 shows evidence of a Fe K emission 
line at 6.4 keV. Moreover, in both observations, the source spectrum exhibits 
an emission feature around 0.7 keV, which is interpreted as due to the
superposition of the K edge absorption features of O and Ne elements with 
uncommon relative abundances with respect to the solar one (O/O$_{\odot} \sim 0.3$, 
Ne/Ne$_{\odot} \sim 2.5$).
In the spectrum of X1556--605 no emission lines are observed.
 We discuss these results and their implications 
 for the source classification and the accretion geometry of the compact object. 
}

\maketitle
\markboth{Farinelli et al.: \sax\ observations of X1543--624 and
X1556--605}{}

\begin{table*}
%\vskip -2cm
%\vskip -4cm
\caption[]{Log of the four \sax\ NFI observations presented in
this work}
\begin{center}
\begin{tabular}{cclccccccc}
\noalign{\smallskip}
\hline
Source & Obs.& \multicolumn{1}{c}{Start Time (UT)} & End Time (UT) &
LECS &  MECS& HPGSPC & PDS & 2--10~keV MECS\\
 & & & &ksec &ksec &ksec &ksec  & count rate (s$^{-1}$)\\
\noalign{\smallskip}
\hline
\noalign{\smallskip}
X1543--624 & 1 & 1997 Feb 21 09:38:02 & 1997 Feb 21 17:39:50  & 6.6 & 16.6 & 7.5 & 7.5 & 12.8\\
            & 2 & 1997 Apr ~~1 22:38:29  & 1997 Apr ~~2 05:54:05  & 6.6 & 18.3 & 7.8 
& 8.1 &11.4\\
\noalign{\smallskip}
\noalign{\smallskip}
\hline
\noalign{\smallskip}
\noalign{\smallskip}
X1556--605 & 1 & 1997 Mar 10 17:10:02  & 1997 Mar 10 23:58:50 &7.3 & 17.9 &7.9 &7.9 
         & 5.42\\
  & 2 & 1997 Apr ~~3 18:08:47  & 1997 Apr ~~4 01:54:50 & 5.1 & 17.8 & 7.3 
& 7.7 & 5.69\\
\noalign{\smallskip}
\hline
\noalign{\vskip 0cm}
\label{t:log}
\end{tabular}
\end{center}
\end{table*}

\section{Introduction}
\label{s:intro}

Low Mass X--ray Binaries (LMXBs) form the most populated class of galactic
X--ray binaries. These objects are formed by late--type
secondary stars, with mass typically less than 1 $M_\odot$, which transfers 
matter 
onto a highly compact primary via Roche lobe overflow. 
These systems mostly contain old, low magnetic field ($<$10$^{9}$ G) 
neutron stars (NS) as compact primaries, and are characterized by persistent, 
albeit variable, X--ray emission. 
The accreting matter carries large angular momentum and this implies the 
formation
of an accretion disk around the compact object; then, X--ray
emission mainly arises from the inner parts of the disk and around the 
accreting
primary, in the so--called boundary layer.
This mechanism ensures quite high efficiency in the conversion of
gravitational energy into X--rays, and luminosities above
10$^{36}$ erg s$^{-1}$ are quite common among LMXBs.

The most successful classification of LMXBs comes from their X--ray `colour' 
behaviour \cite{Hasinger89}. On the basis of this classification, LMXBs are subdivided
into Z sources  and atoll sources, from the shape of their track
in the X--ray colour--colour diagram (CD) and on the different 
timing behaviour that correlates with the position on the tracks. The
observed time scales taken  to track their CD are shorter
(hours or days) for Z sources than for atoll sources (weeks or months).
LMXBs belonging to the atoll sources  usually have luminosities 
lower than Z sources,
typically in the range 0.01--0.1$\times L_{\rm Edd}$.
They are observed in either soft (`banana') or hard ('island') 
state: in the latter case they show, above 10~keV, power--law ({\sc pl}) spectral 
shapes with typical high--energy cut--offs $E_c \la$ 100 keV, except a few 
peculiar cases, like Aql X--1 \cite{Harmon96} and 4U 0614+09 \cite{Piraino99}; the contribution 
of the high energy component with respect to the total flux  can be  
very high
(e.g., about 50\% of the total flux in the case of 4U1705-44, Barret\ et al. 
1996\nocite{Barret96}). It has been demonstrated 
(e.g., van der Klis et al. 1990\nocite{Vanderklis90},
van der Klis 1995\nocite{Vanderklis95}) that the source state along
the CD track correlates well with the mass accretion rate $\dot{M}$, and
the hard state (island) occurs preferrably at low $\dot{M}$.

Spectra of Z sources are generally much softer. In the eighties they
were usually described in terms of a blackbody ({\sc bb}) plus an
additional component: either an unsaturated Comptonization spectrum 
(the 'Western model',
White et al.\ 1986, 1988\nocite{White86,White88}), approximated
by the function $I(E) \propto E^{-\Gamma} exp[-E/kT_e]$ ({\sc usc}) or the 
 solution of the Kompaneets equation given by Sunyaev and Titarchuk (1980)\nocite{Sunyaev80}, or a multi--colour  disk blackbody ({\sc dbb}, the 'Eastern model', 
Mitsuda et al. 1984\nocite{Mitsuda84}).
Later Mitsuda et al. (1989)\nocite{Mitsuda89} refined the Eastern model 
by replacing
the \bb\ with a Comptonized \bb\ and successfully applied it to 
the atoll source X1608--522. 
In the 1--10/20 keV energy range, the classical Western and Eastern models 
have also been used
to describe spectra of atoll sources (e.g., White et al. 1988\nocite{White88}, 
Asai et al. 2000\nocite{Asai00}, hereafter A2000). Also other models, like {\sc dbb} or {\sc bb}, 
plus the Comptonization model worked out by Titarchuk (1994)\nocite{Titarchuk94} 
have been adopted to describe spectra of Z sources (e.g., Di Salvo et al. 
2001, 2002\nocite{Disalvo01,Disalvo02}).
Recently it has been found that most of the classical Z sources 
(GX 5--1, Asai et al 1994\nocite{Asai94}; Cyg X--2, Frontera et al. 
1998\nocite{Frontera98}, Di Salvo et al. 2002\nocite{Disalvo02}; 
GX 17+2, Di Salvo et al. 2000a\nocite{Disalvo00a}; GX 349+2, Di Salvo
et al. 2001\nocite{Disalvo01}, Sco X--1, D'Amico et al. 2001\nocite{Damico01}), 
and Cir X-1 \cite{Iaria01} show in their spectra a hard
X--ray tail. The tail
is observed only during some positions of the sources along their 
CD, with no general rule: e.g., in GX17+2 the hard tail is apparent
during the Horizontal Branch, in GX349+2 it is detected during
the Flaring Branch, in Sco X--1 it is detected in all the three branches 
of the Z pattern.
 
It is still not clear what determines the presence of the
high energy component in Z sources and whether this component evolves with
continuity from atoll sources to Z sources. Recently Gierli\'nski \& Done (2002), 
analyzing
{\it RXTE} observations of three atoll sources (Aql X-1, 4U 1608-52 and  
4U 1705-44), and Muno et al. (2002), analyzing a sample of 15 LMXBs, showed
that atoll sources that exhibit X--ray intensity 
variations by more than a factor 10 trace  a Z--pattern in the 
CD like the Z sources and concluded that these atoll sources evolve
with the mass accretion rate in similar ways to
the Z sources. According to Gierli\'nski \& Done (2002) \nocite{Gierlinski02}, 
both sets of sources undergo a transition from the upper to the lower branch of the
CD when the disk, which is truncated during the upper
branch, penetrates down to the NS surface, but in the case of
the atolls, the disk truncation is due to mass evaporation while in
the case of the Z sources this is due  to the NS magnetic field. However,
as also pointed out by Muno et al. (2002), while the time evolution of both 
classes of sources could be similar, their spectral behaviour is notably
different, making the unification of the two classes  of sources difficult. 
Also, both studies did not address the timing
information associated with the motion through the CD, so that their
conclusion that a Z--shape is traced out is not yet entirely secure.

Among LMXBs there are still unclassified sources: on the basis of
the data already available, these sources do not display any peculiarity
in their X--ray light curve or X--ray colour--colour behaviour.
\sax\ \cite{Boella97a} offers the possibility to investigate them, 
thanks to the broad energy band (0.1--200 keV) and high
sensitivity of the Narrow Field Instruments (NFIs) onboard.
We thus started an observational campaign on a sample of these
LMXBs. In this paper we report results for X1543--624 and X1556--605.
The paper is organized as follows: in Section 2 we report the present status
of knowledge of X1543--624 and X1556--605, in Section 3 we describe our
observations and the  data analysis, in Section 4 we present the results,
in Section 5 we discuss them and in Section 6 we draw our conclusions.

\section{The X--ray sources X1543--624 and X1556--605}

\subsection{X1543--624}
\label{X1543--624:l}
X1543--624 was classified by Warwick et al. (1981) \nocite{Warwick81}
as a persistent LMXB with a  2--10~keV mean flux
of $\sim$ 7$\times 10^{-10}$ erg cm$^{-2}$ s$^{-1}$,  which can vary by 
a factor $\sim$ 2. Apparao et al. (1978)\nocite{Apparao78} provided an
accurate position of the source (error radius of 30$''$) with 
a list of candidate optical counterparts.
Subsequently McClintock et al. (1978), by means of spectrophotometric
measurements, found that the most likely counterpart was the
object \#6  of Apparao et al. (1978) (magnitude $B\ga$ 20), which shows a
very blue colour. The corresponding  X--ray to optical luminosity is   
$L_{\rm X}$/$L_{\rm opt} \sim 1.4 \times $10$^3$ \cite{Bradt83}, a value
quite typical for LMXBs \cite{Jvp95}.
Smith et al. (1990)\nocite{Smith90} with IRAS discovered a far--infrared
counterpart of
X1543--624  having a spectral slope compatible with that of an accretion
disk. No radio counterpart of the source was detected \cite{Wendker95}.

Singh et al. (1994, hereafter S94), using {\it EXOSAT}
archive data, fitted the 1--20 keV source spectrum with an absorbed
{\sc bb} plus a Comptonization model ({\sc compst}, Sunyaev \& Titarchuk
1980).
This fit (albeit it poorly constrained the parameters of the
{\sc compst} model)
yielded an hydrogen column density of $\sim 1.3 \times $10$^{22}$~cm$^{-2}$,
a \bb\ temperature $kT_{\rm bb} \sim $ 1.6 keV and a Comptonizing cloud
with electron temperature $kT_{\rm e} \sim $ 37 keV (lower limit 4~keV)
and optical depth $\tau \sim$ 1.2 (upper limit 4.8).
These authors also found a broad iron emission line at $\sim$7~keV 
 with an Equivalent Width (EW) of $\sim$ 100 eV. 

Christian \& Swank (1997), using {\it Einstein}
archival data between 0.5 and 20~keV, found that the best fit,
 even if unsatisfactory, of the source photon spectrum was obtained with either 
an {\sc usc} model with  
$\Gamma \sim 1.9$ and $\kte \sim 25$ keV  or a {\sc bb} plus thermal
bremsstrahlung ({\sc tb}) model, with $kT_{\rm bb} \sim $ 2 keV 
and $kT_{\rm tb} \sim $ 2.4 keV, both models being photoelectrically--absorbed.

Simpler models, such as a \bb\ or a \pl\ or a
{\sc dbb}, provided worse fits.
The estimated hydrogen column density was $N_{\rm H} \sim 2.5 \times
10^{21}$~cm$^{-2}$, a value about 4 times lower than that found by S94,
while the source unabsorbed flux was $\sim$ 1$\times
10^{-9}$~erg cm$^{-2}$ s$^{-1}$. No evidence of emission lines was found.

More recently A2000, in the framework of an
archival survey of iron K lines in LMXBs,
analyzed an observation of this source performed with ASCA on August 17 1995.
During the observation the 1--10~keV unabsorbed flux of the source 
was $1.1 \times 10^{-9}$ erg~cm$^{-2}$ s$^{-1}$, which is consistent with
that found by Christian \& Swank (1997).
The 0.7--10~keV source spectrum was fit with an absorbed {\sc bb} plus
{\sc dbb} model with
$kT_{\rm bb} \sim 1.6$ keV,  a temperature of the inner disk $kT_{\rm in} 
 \sim 0.7$ keV and $N_{\rm H} \sim 1.4\times 10^{21}$~cm$^{-2}$, which is
consistent with the estimate by Christian \& Swank (1997). A2000
\nocite{Asai00} also
marginally detected an iron emission line at 6.8~keV with EW $\sim$ 48 eV.
A low energy emission feature near 0.7 keV was observed
from this source (White et al. 1997).
Juett et al. (2001, hereafter J2001), stimulated by the results obtained 
on 4U 0614+091 with {\it Chandra}, reanalyzed the {\it ASCA} data
of X1543--624 adopting a \bb\ plus \pl\ model 
photoelectrically--absorbed by a column density  $N_{\rm H}$ with
 Ne and O  abundances free to vary in the fit. 
The model, even if unsatisfactory (\chiq/dof = 1542/754), was
suitable to describe the emission feature at 0.7 keV 
with Ne/Ne$_{\odot} \sim 2.9$ and O/O$_{\odot} \sim 0.5$.
Very recently Schultz (2002) analyzed archival {\it ASCA}, \sax\ and {\it RXTE} 
data obtained from the observations of X1543--624. For \sax\ only LECS and MECS
data were considered, while for {\it RXTE} only PCA data were analyzed. 
The spectra of each of the observations were fitted  with an absorbed \bb\ plus 
\compst\ model. The  2--10 keV source luminosity was observed to increase by a factor
$\sim 1.6$ (from $8.6 \times 10^{36}$~erg~s$^{-1}$ to  $ \sim 1.4 \times 
10^{37}$~erg~s$^{-1}$), with harder spectra ($\kte$ from $\sim 0.5$ to $\sim 3.5$ keV
and $\tau$ from $\sim 30$ to $\tau \sim 10$ ) at higher luminosities, unlike 
the general behaviour of LMXB sources (see, e.g., Bloser et al. 
2000\nocite{Bloser00}). Evidence of an Fe K emission line in the higher luminosity 
spectra obtained during the {\it RXTE} observations was reported.

\subsection{X1556--605}
\label{X1556--605:literature}
X1556--605 was classified by Warwick et al. (1981) \nocite{Warwick81}
as a persistent, irregularly variable LMXB with a 2--10~keV flux of
$\sim 3 \times 10^{-10}$~erg~cm$^{-2}$~s$^{-1}$, with variations within a
factor $\sim$ 4 \cite{Bradt83}.
%The source was detected in X--rays by several satellites.
A precise celestial position (error radius of 30$''$)
%($\alpha_{1950}$ = 15$^{\rm h}$ 56$^{\rm m}$ 47$\fs$2;
%$\delta_{1950}$ = --60$^{\circ}$ 35$'$ 47$''$,
was given by Apparao et al. (1978) using a {\it SAS-3} observation
made on June 1975. This position allowed Charles et al. (1979) to identify
the optical counterpart of the source with the star \#43 (also known as LU TrA)
in the finding chart reported by Apparao et al. (1978) on the basis of its
spectrophotometric characteristics. This object is rather
blue ($U$--$B$ = --0.7) and displays a He {\sc ii} $\lambda$4686 line in
emission. These features are quite common in the optical counterparts of
LMXBs. The optical counterpart showed a magnitude $V \sim
19$ and H$_\alpha$, He {\sc ii} $\lambda$4686 and N {\sc iii}
$\lambda$4640 (Bowen blend) emission lines (Motch et al. 1989, hereafter M89).
 These
authors also found optical variability of $\sim$ 0.4 mag, not correlated
with the X-ray emission, on timescales of hours.
Using the data of M89, the X-ray/optical
luminosity ratio
of the source was $\sim$ 1200 (and not 180 as erroneously
reported by these authors), in better agreement with the value,
2$\times $10$^3$, given by Bradt \& McClintock (1983).
A possible photometric orbital period of 9.1 hours, a value not
inconsistent with the size and characteristics of a LMXB, was found 
by Smale (1991) from optical $V$--band observations.
Lastly, this source was also detected in the far--infrared with IRAS
\cite{Smith90}, while no counterpart was found at radio wavelengths
\cite{Wendker95}.

The {\it Einstein} 0.5--20~keV spectral data \cite{Christian97} were 
better described with either an absorbed \usc\ model with 
$N_{\rm H} \sim 3.7\times 10^{21}$~cm$^{-2}$, photon index 
$\Gamma \sim 0.58$ and $\kte \sim3.3$~keV or with a \bb\ plus {\sc tb} model with 
$N_{\rm H} \sim 4.6 \times 10^{21}$~cm$^{-2}$, 
$kT_{\rm bb} \sim 1.6$~keV and $kT_{\rm tb}\sim$ 4.9~keV, even though the fits were 
not completely satisfactory.
M89 performed a 
quasi--simultaneous X--ray/optical
observational campaign on this source during the years 1984/1985.
The X--ray data, collected with {\it EXOSAT}, showed flux
variations by $\sim$ 20\% accompanied by changes in the hardness 
ratio of the emission, but did not show bursts or pulsations.
 The photoelectrically--absorbed 0.1--10 keV X--ray spectrum was fit 
 with either a
\bb\ plus {\sc tb} model ($N_{\rm H} \sim 4\times 10^{21}$~cm$^{-2}$, 
$kT_{\rm bb} \sim$ 1.3 keV, $kT_{\rm tb} \sim$ 7 keV) or with an {\sc usc} ($N_{\rm H} \sim 3 \times  10^{21}$ cm$^{-2}$, 
 $\Gamma \sim 0.7$, $\kte \sim 3.8$ keV), 
 or with a {\sc compst} model ($N_{\rm H} \sim 5 \times 10^{21}$~cm$^{-2}$,
 $kT_e \sim$ 2 keV, $\tau \sim$ 20).

Notice that the estimated  column density is consistent with that derived
with the {\it Einstein} data.
The source fluxes were comparable (1--10~keV flux of $\sim 4 \times 10^{-10}$~erg~cm$^{-2}$ s$^{-1}$) during the EXOSAT and the {\it Einstein} observations.
No Fe K emission line   was detected in the spectrum (see also
Gottwald et al. 1995).

\section{Observations and data analysis}
\label{s:obs}

\begin{figure*}
\begin{center}
\epsfig{figure=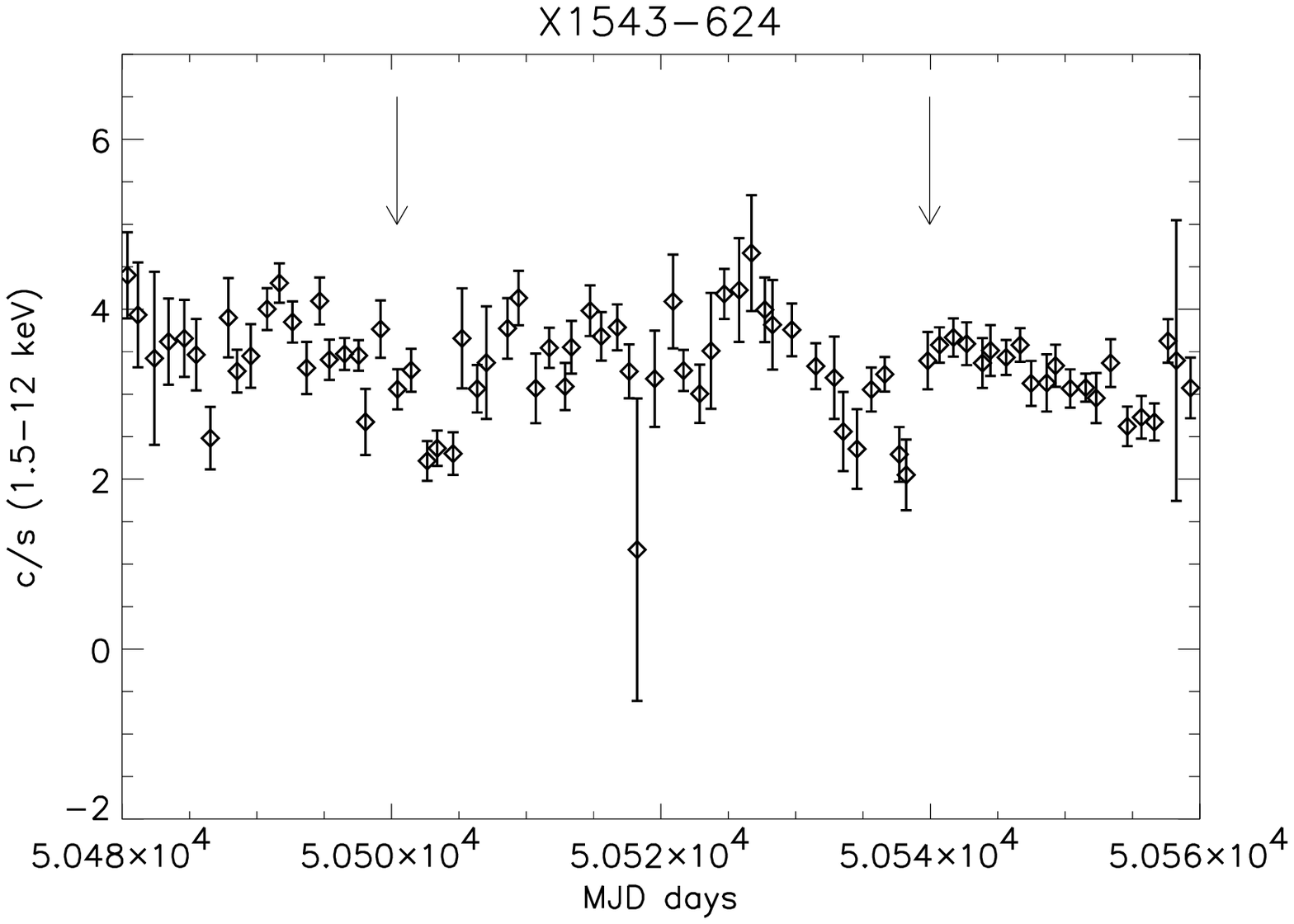,width=9cm}
\epsfig{figure=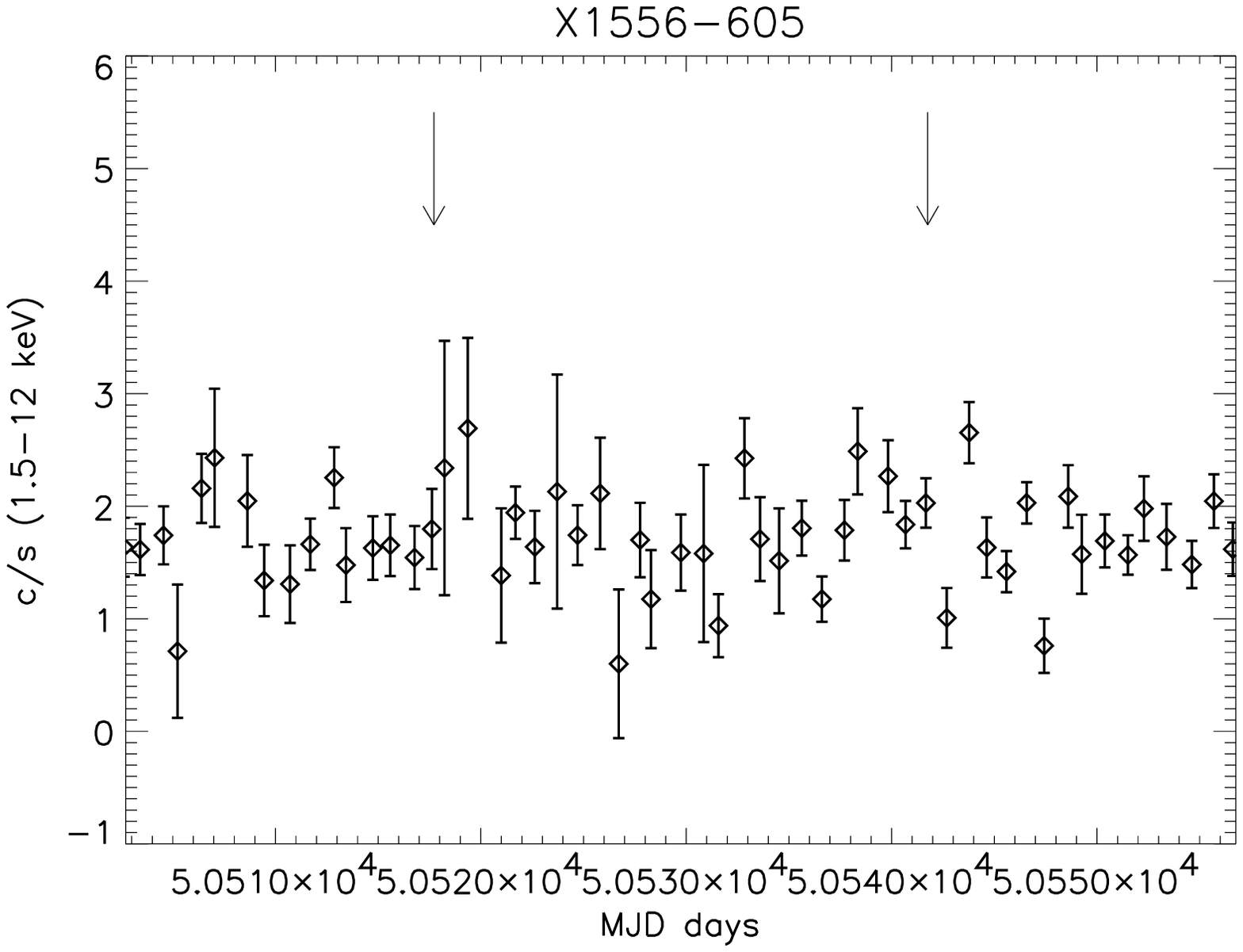,width=9cm}
\end{center}  
\vspace{-0.1cm}
\caption[]{The one day average light curves of X1543--624 and X1556--605 
detected with the {\it RXTE} ASM in the 1.5--12 keV energy band. The 
arrows mark the epoch of the \sax observations of each source. 
ASM data can be retrieved on the public archive at 
{\it http://xte.mit.edu/XTE/asmlc/ASM.html}.}
\label{f:asm}
\end{figure*}

X1543--624 and  X1556--605 are continuously  monitored
with the {\it All Sky Monitor} (ASM) onboard RXTE. The ASM 
light curve of both sources in the  February--April 1997 period which
covers our observations is shown in Fig.~\ref{f:asm}. We observed the
sources twice (see log in Table~\ref{t:log}) with the \sax NFIs. 

These include a Low--Energy Concentrator Spectrometer (LECS,
0.1--10~keV; Parmar et al. 1997), three Medium--Energy Concentrator
Spectrometers (MECS, 1.5--10~keV; Boella et al. 1997b), a High Pressure
Gas Scintillation Proportional Counter (HPGSPC, 4--120~keV; Manzo et al.
1997), and a Phoswich Detection System (PDS, 15--300~keV; Frontera et
al. 1997).
At the time of these observations, Unit 1 of MECS was still operative (it
failed shortly thereafter, on May 6 1997), so the four pointings 
reported here were performed with all three MECS units.
Table~\ref{t:log} gives the observation log along with the exposure times.
During all pointings the  NFIs worked nominally and both sources
were detected  in all of them.

Good data were selected from intervals when the NFIs elevation angle was
above the Earth limb by at least $5^{\circ}$ and, for LECS, during
spacecraft night time.
The SAXDAS 2.0.0 data analysis package \cite{Lammers97} was used for 
the processing of the LECS, MECS and HPGSPC data. The PDS data reduction was
 performed using the XAS package v2.1 (Chiappetti \& Dal Fiume 1997).
LECS and MECS spectra were extracted from a region of 8\arcmin\ radius
centred on the source position after proper background subtraction.
 For LECS and MECS we used the background measured pointing a blank--field 
\cite{Fiore99}, while for  HPGSPC and PDS  we continuously monitored it
using the rocking--collimator technique \cite{Frontera97}.  
The rocking angle and attitude was suitably choosen in order to
point to sky fields with no known X--ray sources in the field of view of the 
instruments.
The  spectra  were rebinned leaving an oversample by a factor of 3 of
 the energy resolution (FWHM), and having a minimum of 20 counts in
each bin such that the $\chi^2$ statistics could reliably be used.
Data were selected in the energy ranges where 
the instrument responses are better known: 0.4--4.0~keV for the LECS,
1.8--10~keV for the MECS, 8--30~keV for the HPGSPC, and 15--200~keV for
the PDS. We  used the package \xspec\ v11.0.1 \cite{Arnaud96} to
fit the multi--instrument spectra.
In the broad--band fits, normalization factors were applied to LECS, HPGSPC 
and PDS spectra following the cross--calibration tests between these
instruments and the MECS \cite{Fiore99}.
The normalization factor of PDS was fixed to 0.9 for both
 sources. An a--posteriori check showed that indeed the best fit model parameters did
 not undergo any significant variation changing this factor in
the range from 0.75 to 0.98 as prescribed by Fiore et al. (1999)\nocite{Fiore99}.
Photoelectric absorption was modeled using the cross sections implemented
in {\sc xspec} \cite{Morrison83} and, when the element abundances were
fixed, we used the standard values given by
Anders \& Grevesse (1989)\nocite{Anders89}. Finally, we assumed a distance
$d$ = 10~kpc for both sources, as done in previous works.
Actually, Christian \& Swank (1997)
gave a  value of 4~kpc for the distance to X1556--605 on the basis
of the optical photometric period measured by Smale (1991) \nocite{Smale91}. 
However, the period inferred by this author lacks confirmation. For
this reason we preferred to assume the ``standard'' value of
10~kpc also for X1556--605.
Uncertainties in the parameters obtained from the spectral fits are single
parameter errors at a 90\% confidence level. The values quoted in square parentheses 
in Table \ref{t:results} are kept frozen during the fit.

%Figure 1

% Figure 2
%
\begin{figure*}
\begin{center}
\epsfig{figure=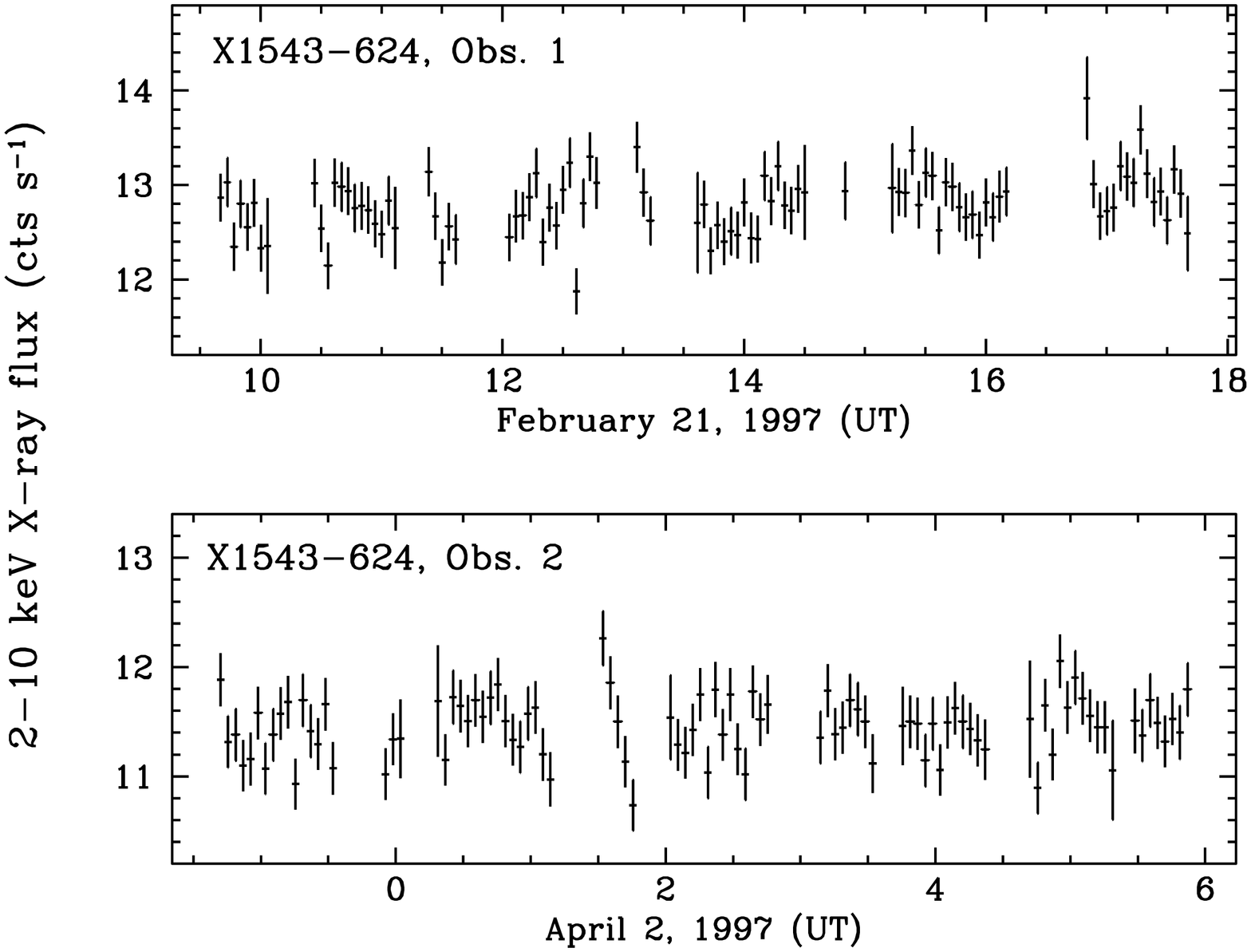,width=10cm,height=16cm, angle=-90}
\epsfig{figure=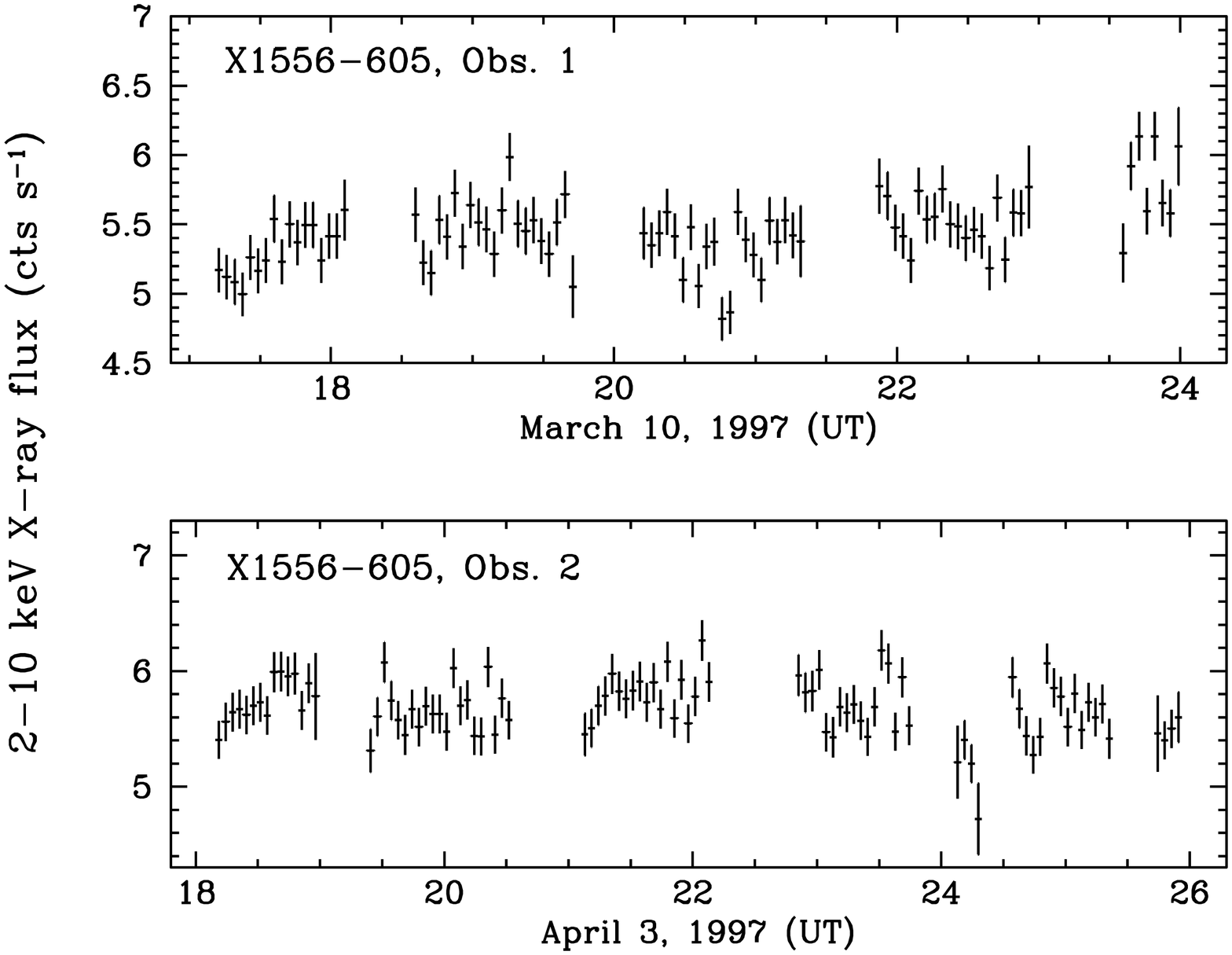,width=10cm,height=16cm, angle=-90}
\end{center}  
\vspace{-0.1cm}
\caption[]{{\it Upper panels:}~2--10~keV MECS light curves of the two \sax\
 observations of X1543--624.
{\it Bottom panels:}~2--10~keV MECS light curves of X1556--605.
Times are expressed in UT of the day reported
in each panel. The time binning is 200~s.}
\label{f:lc}
\end{figure*}

%
% Figure 3
%
\begin{figure*}
\begin{center}
\epsfig{figure=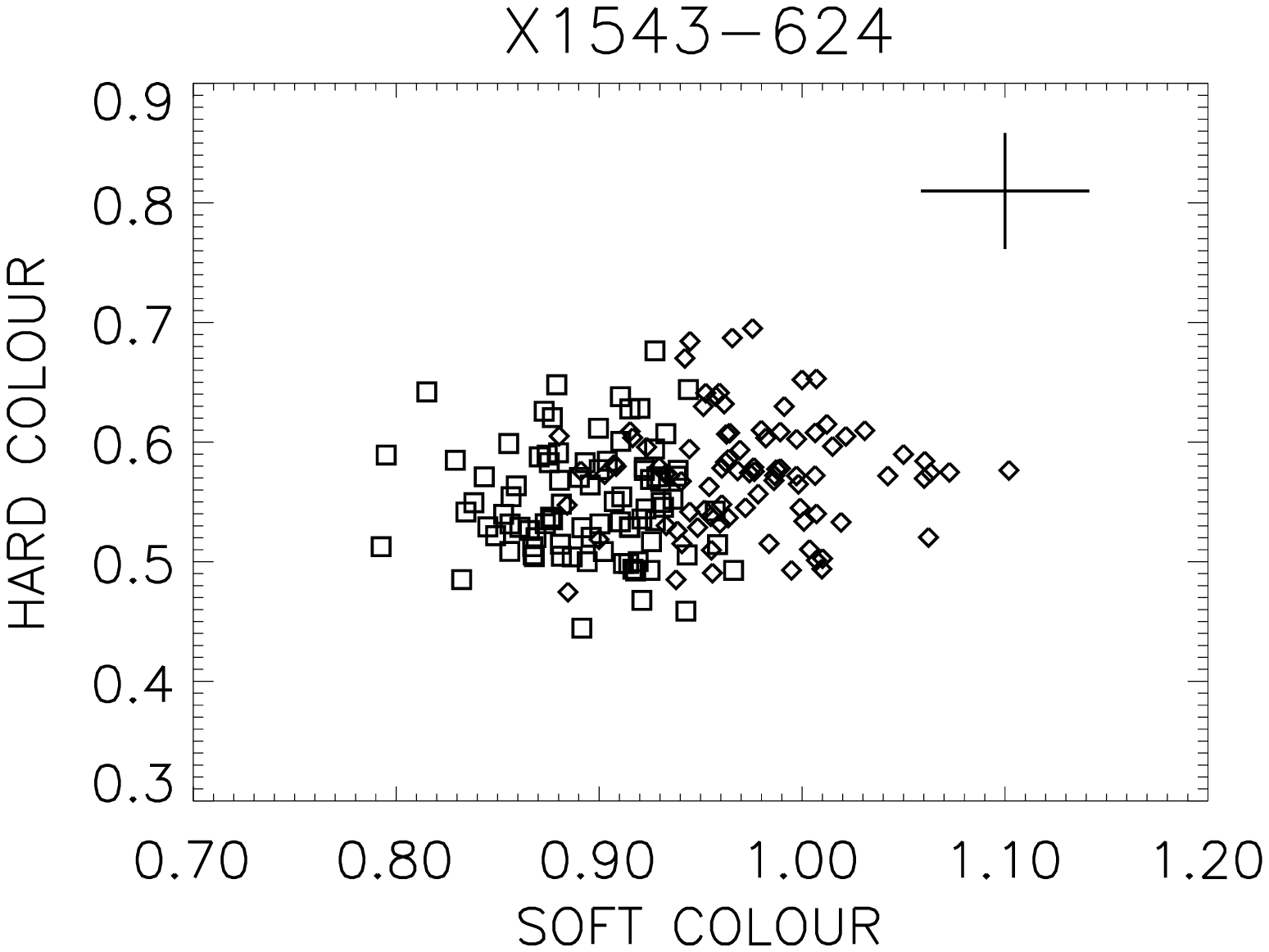,height=6cm, width=6.5cm}
\epsfig{figure=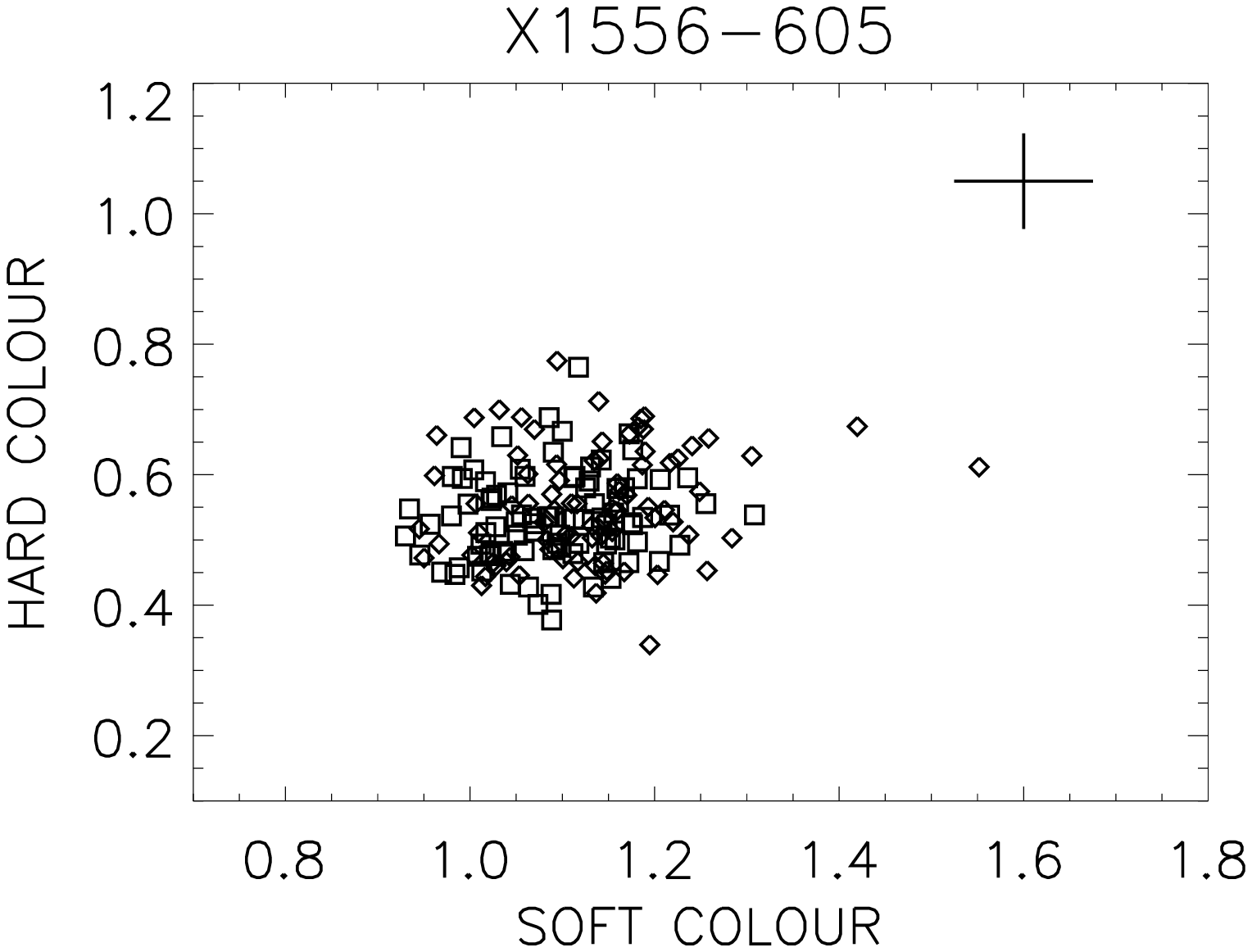,height=6cm, width=6.5cm}
\end{center} 
%\vspace{-6cm}
\caption[]{Colour--colour diagrams of both observations of X1543--624 ({\it
left panel}) and X1556--605 ({\it right panel}). The y axis ({\it hard colour})
gives the ratio  between the 6.5--10 keV and 5--6.5 keV count rates. The
x axis ({\it soft colour}) gives the count rate ratio in the 3--5 keV
and 1.8--3~keV energy bands. In
both diagrams each point corresponds to a time interval of 200 s. Typical 
error bars of the two colours are reported in the top right of each panel.
{\it Diamonds}: first observation; {\it squares}: second observation.} 
\label{f:cd}
\end{figure*}

\section{Results}
\label{s:results}
Figure~\ref{f:lc} shows the 2--10 keV light curves of both sources during
the two observations. Neither dips nor eclipses nor X--ray bursts are
observed.
On time scales of several hundreds of seconds, there is evidence of a
moderate time variability from X1543--623, while a small flux decrease
(by $\sim$ 9\% in the case of X1543--623 and by $\sim$ 5\% in the case
of X1556--605) is measured from the first to the second observation.

The MECS light curves in four energy ranges, namely 1.8--3 keV,
3--5 keV, 5--6.5 keV, 6.5--10 keV,  were extracted to  study the
CD. The soft colour was obtained from
the 3--5 keV/1.8--3 keV count rate ratio, the hard colour from the
6.5--10 keV/5--6.5 keV ratio. The CDs of both sources
 are shown in Fig.~\ref{f:cd}. As can be seen, both the soft and
hard colours of X1543--623 and X1556--605 change by about 20\% during the
observations, with an almost circular pattern traced by the sources.
In the case of X1543--623
the soft colour (x axis) has a centroid value of $\sim 0.95$ and the 
hard colour (y axis) of $\sim 0.55$, in the case of X1556--605 the hard colour
has a centroid value similar to that of X1543--624, while the soft colour
has  a higher centroid ($\sim$1.1).

\subsection{Spectral properties}
\label{s:spectra}

As already said in Section~\ref{s:intro}, several models have been proposed 
to fit the spectra of both Z and atoll sources.
In the cases of low luminosity sources, a single component model has sometimes
provided good results (e.g., White et al. 1988\nocite{White88}).
We thus performed our analysis first with a single component model and
then with two--component models both in their ``classical'' forms
\cite{White86,White88,Mitsuda84,Mitsuda89} and in the forms
 proposed more recently (see references in Section \ref{s:intro}).

\subsubsection{X1543--624}
\label{ss:1543}

\begin{figure*}
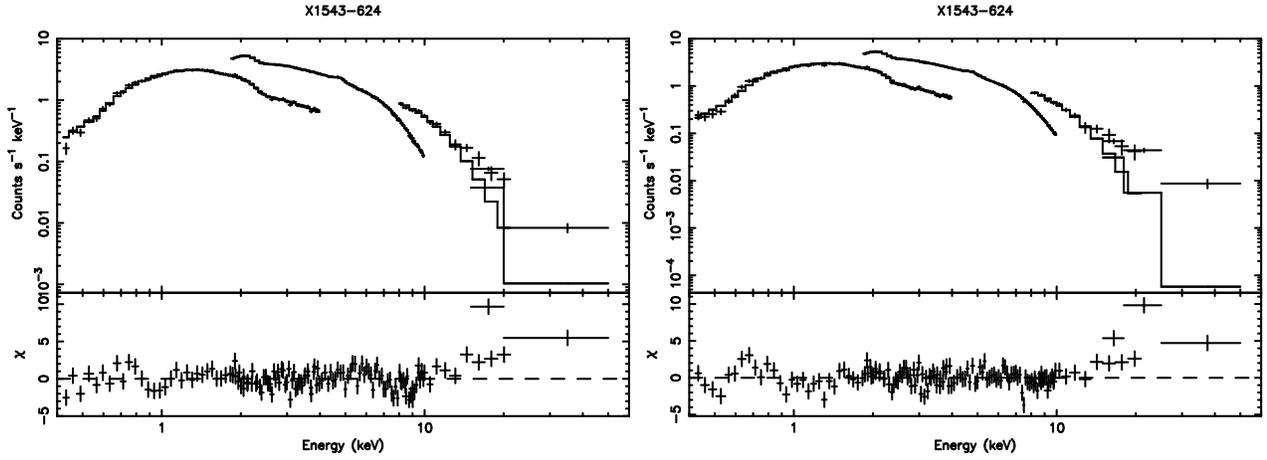

\begin{center}
\epsfig{figure=ms2663_f7.ps,width=6cm,angle=270}
\epsfig{figure=ms2663_f8.ps,width=6cm,angle=270}
\end{center}
\caption[]{Count rate spectra of the two observations of X1543--624
along with the folded model \bb\ plus \dbb, photoelectrically--absorbed
(\wabs)  and residuals to the model in units
of $\sigma$. 
{\it Left panel}: first observation. {\it Right panel}: second observation. 
The excess above 15 keV can be recovered
replacing the \bb\ with a Comptonized \bb\ (see text).}
\label{folded_bbdiskbb_x1543}
\end{figure*}

\begin{figure*}
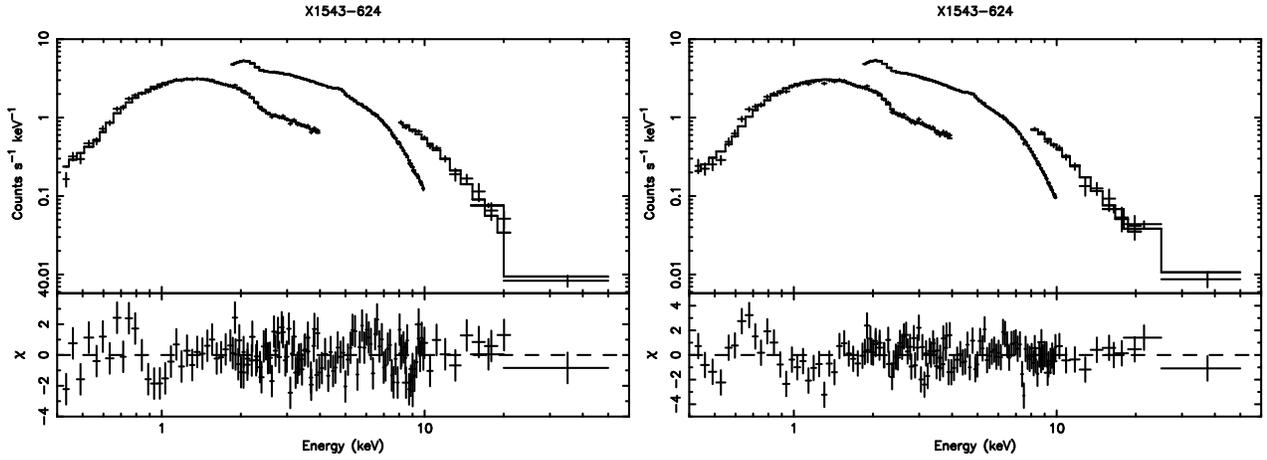

\begin{center}
\epsfig{figure=ms2663_f9.ps,width=6cm,angle=270}
\epsfig{figure=ms2663_f10.ps,width=6cm,angle=270}
\end{center}
\caption[]{Count rate spectra of the two observations of X1543--624
along with the folded model \dbb\ plus \compbb, photoelectrically--absorbed
(\wabs)  and residuals to the model in units
of $\sigma$. 
{\it Left panel}: first observation. {\it Right panel}: second observation. 
}
\label{folded_diskbbcompbb_x1543}
\end{figure*}

 We first tried photoelectrically--absorbed (\wabs\ in {\sc xspec}) single component models: 
a simple {\sc pl}, an \usc\ and the Comptonization models \compst\ and   \comptt. 
None of them was found to
provide a satisfactory description of the data  ($\chi^2_{\nu}$ always higher than 
4). Thus we concentrated on two--component models.
 The \bb\ plus \usc\ model provided  \chiq/dof = 288/133 
and 367/133 in the first and second observation, respectively, with
residuals to the model sinusoidally  distributed along the entire energy
band. Also the 
{\sc bb} plus {\sc dbb} did not give a good description of the data (\chiq/dof = 
352/134  and 335/134 for the first and second observation, respectively), but,
 in this case, apart from an excess count around 0.7 keV, the residuals 
 are mainly concentrated  above 15 keV (see Fig.~\ref{folded_bbdiskbb_x1543}).
 Replacing  the simple \bb\ with 
 a Comptonized \bb\  (\compbb\ in {\sc xspec}; Nishimura et al. 1986) we obtained
a much better description of the data (\chiq/dof = 180/132 for the first observation and
\chiq/dof = 175/132 for the second one), with the residuals to the model
 only  at about  0.7 keV (see Fig. \ref{folded_diskbbcompbb_x1543}).
 Other two-component models used for LMXBs, like  a \bb\ or \dbb\ plus a 
 {\sc comptt}, did not provide better fits to the data.

 As noticed above, irrespective of the adopted model, positive residuals are
 present in both observations around 0.7 keV. 
Assuming as input model the \bb\ plus \compbb, which gives the best fit 
to the
continuum spectrum, these residuals can be fit either 
adding a Gaussian (\chiq/dof = 161/129 and 129/129 in the first and second 
observation, respectively), or, following J2001, assuming an 
 absorption with free abundances of O/O$_\odot$ and Ne/Ne$_\odot$ (\vphabs\
 model in {\sc XSPEC}, \chiq/dof = 160/130 and 131/130, in the first and second
observation, respectively).
The Gaussian centroid energy ($ \sim 0.65$ keV) is   
consistent with fluorescence emission from O{\sc vii} or O{\sc viii}, and
the abundances of O and Ne, assuming the \vphabs\ model, are reported in 
Table~\ref{t:results}. For the reasons 
discussed by J2001 we prefer the interpretation of the 0.7 keV excess in terms of 
absorption in neon--rich material local to the binary.

\begin{table*}
\caption[]{Best fit  continuum parameters of the observed LMXBs.} 
\begin{center}
\begin{tabular}{ccc|cc}
\hline
\noalign{\smallskip}
  & \multicolumn{2}{c}{X1543--624}   & \multicolumn{2}{c}{X1556-605}  \\
\noalign{\smallskip}
\hline
\noalign{\smallskip}
Parameter   & Obs. 1             & Obs. 2    &  \multicolumn{2}{c}{Obs. 1+2} \\ 
            & \multicolumn{2}{c}{\vphabs(\dbb\ + \compbb\ + \gaussian)} & \wabs(\dbb\ +\compbb)  & 
              \wabs(\bb\ + \usc) \\
             
\noalign{\smallskip}
\hline
\noalign{\smallskip}

$N_{\rm H}^{a}$ & 0.21$^{+0.04}_{-0.03}$  &  0.31$^{+0.02}_{-0.06}$  & 
        0.30$^{+0.01}_{-0.02}$  &  0.37$^{+0.03}_{-0.06}$  \\

$kT_{\rm bb}$ (keV) & 1.45$^{+0.01}_{-0.01}$  & 1.44$^{+0.01}_{-0.01}$ &  
         1.49$^{+0.03}_{-0.02}$   & 1.49$^{+0.23}_{-0.09}$ \\

$kT_{\rm e}$ (keV)        & 6.7$^{+1.5}_{-0.7}$ & 25.4$^{+5.2}_{-4.3}$  
                    & 17.5$^{+12.6}_{-8.4}$ &  4.1$^{+0.5}_{-0.8}$  \\

$\tau$ 		    & 1.3$^{+0.1}_{-0.1}$ & 0.5$^{+0.1}_{-0.1}$& 0.5$^{+0.2}_{-0.1}$ 
                    & 	--		    \\
$\Gamma$	    &	--		     &	--		     & --
                    & 1.0$^{+0.2}_{-0.3}$ \\

$kT_{\rm in}$ (keV) & 0.64$^{+0.01}_{-0.01}$ & 0.62$^{+0.02}_{-0.02}$ & 
           0.93$^{+0.06}_{-0.05}$   &  -- \\

$R_{\rm in} \sqrt cos($i$)$ or $R_{\rm bb}$ (km) & 14.1$^{+1.0}_{-0.8}$ & 16.2$^{+1.1}_{-1.1}$ & 
             3.9$^{+0.4}_{-0.4}$         &    1.2$^{+0.2}_{-0.5}$ \\

$L_{\rm dbb}^{b}$ or $L_{\rm bb}^{b}$&  8.6 & 9.7 & 2.9 &  1.1 \\

\noalign{\smallskip}

 $E_l$ (keV) & [6.4] & [6.4] & --      & -- \\

 $\sigma_l$  (keV)  & 0.7$^{+0.4}_{-0.4}$  & 0.7$^{+1.1}_{-0.7}$ & --      & -- \\
 
 $I_l$ ($10^{-4}$ cm$^{-2}$ s$^{-1}$)  &  5.3$^{+4.7}_{-3.1}$     & 3.8$^{+10.8}_{-3.1}$ & --      & -- \\

 $EW_l$ (eV) & 65$^{+58}_{-39}$  & 55$^{+159}_{-45}$   &  --      & -- \\

\noalign{\smallskip}

\noalign{\smallskip}

O/O$_{\odot}$  & $0.32^{+0.26}_{-0.27}$  & $<$0.33 & --      & -- \\

Ne/Ne$_{\odot}$ & $2.4^{+0.5}_{-0.7}$ & $2.8^{+1.0}_{-0.6}$  & -- & -- \\

\chiq/dof & 150/129          &  125/129      & 134/124         & 141/124 \\

\noalign{\smallskip}
\hline
\noalign{\smallskip}

$L_{\rm 0.1-200~keV}^{b}$ & 18.9           & 18.6      &  6.8    & 7.2 \\

$L_{\rm 1-20~keV}^{b}$   & 15.0           & 13.8      &  5.8     & 5.9 \\

$L_{\rm 20-200~keV}^{b}$  & 0.3         & 0.6      &   0.14    & 0.05 \\

\noalign{\smallskip}
\hline
\noalign{\smallskip}

\multicolumn{5}{l}{$^{a}$ In units of 10$^{22}$ cm$^{-2}$} \\

\multicolumn{5}{l}{$^{b}$ Unabsorbed luminosity in units of 10$^{36}$ erg 
s$^{-1}$ assuming a distance of 10 kpc for both sources.}\\

\noalign{\vskip -0.cm}
\label{t:results}
\end{tabular}
\end{center}
\end{table*}

\begin{figure*}
\begin{center}
\epsfig{figure=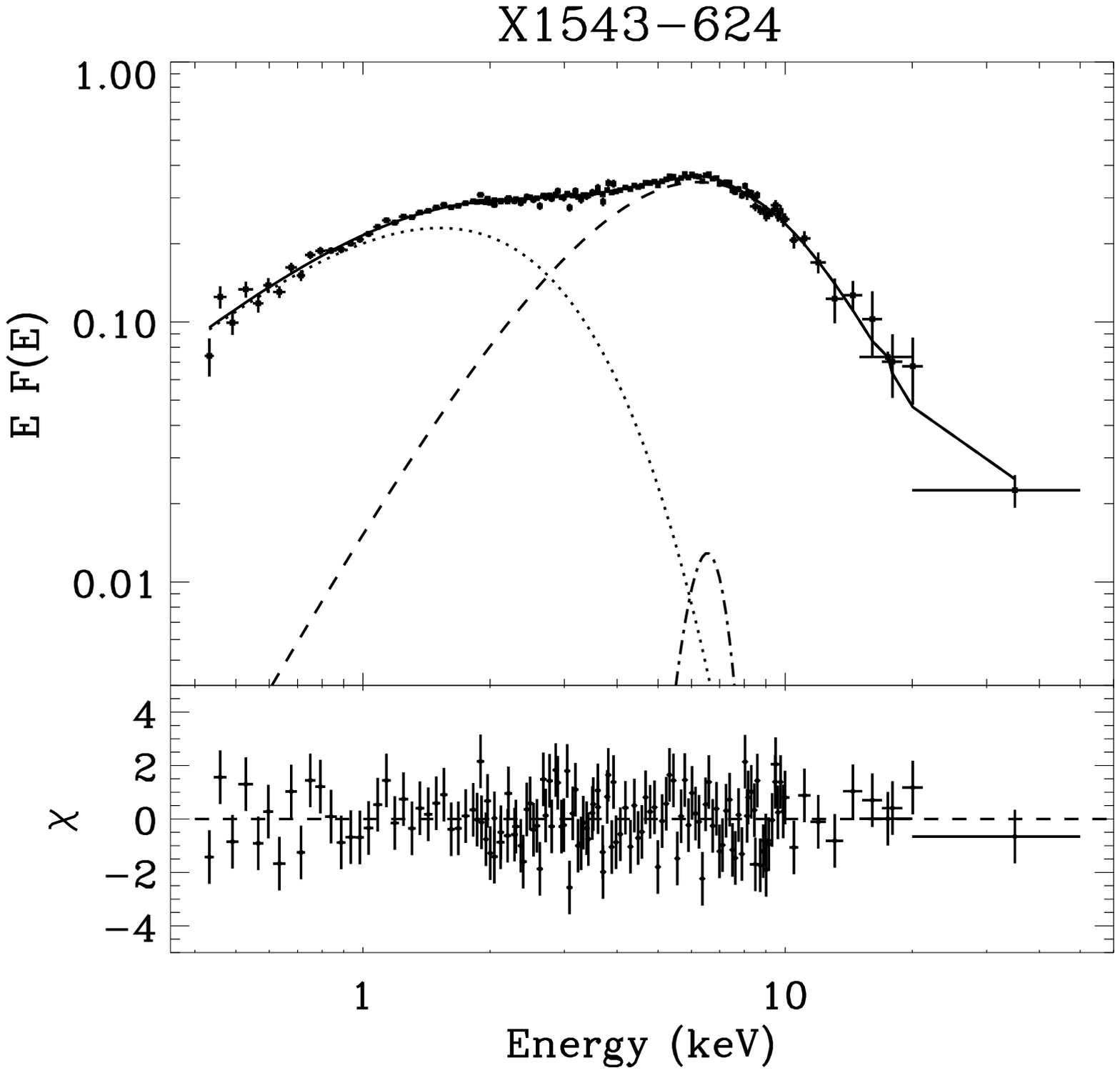,height=6cm,width=6.5cm}
%\vspace{1cm}
\epsfig{figure=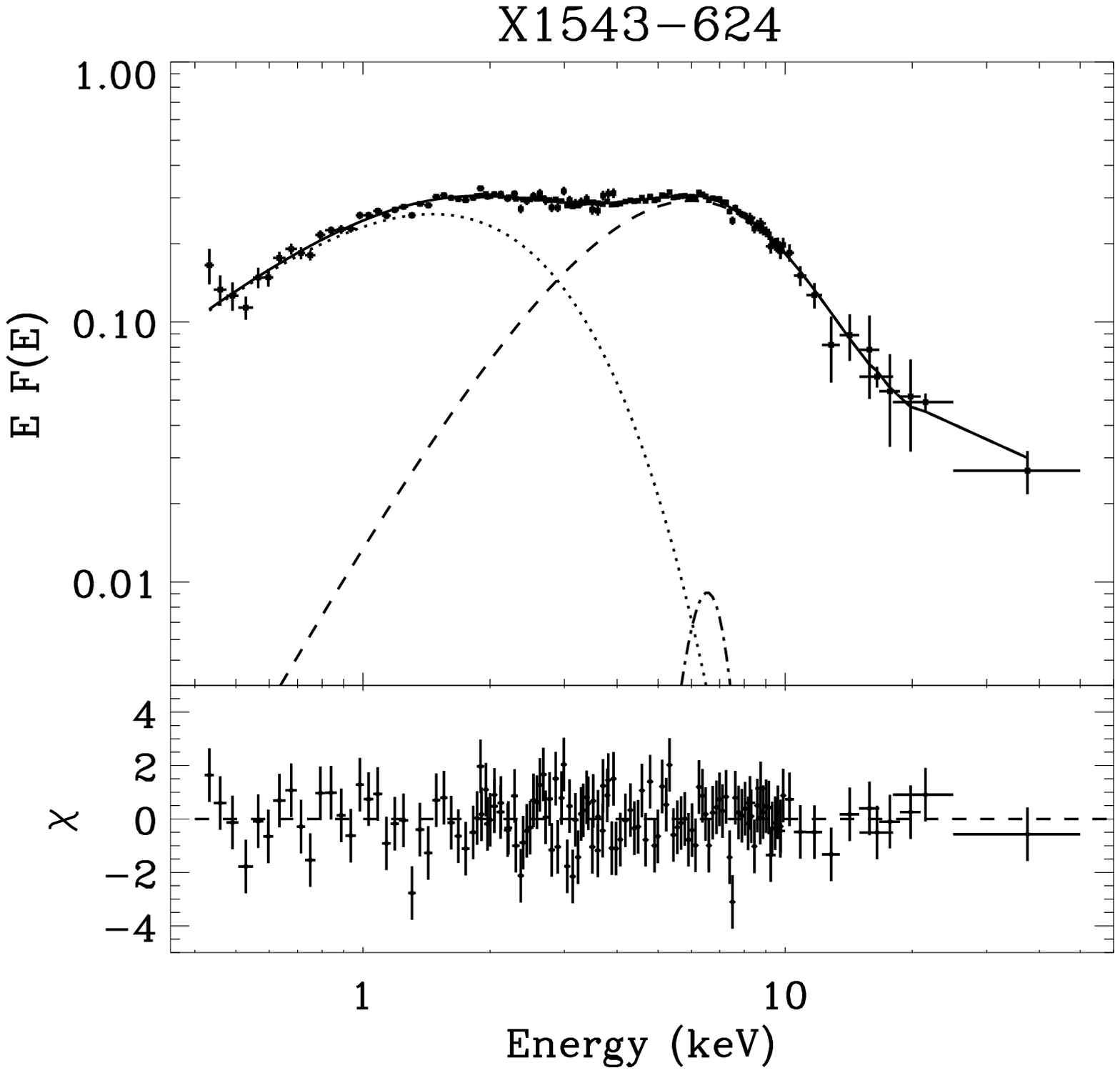,height=6cm,width=6.5cm} 
\vspace{0.5cm}
\caption[]{Absorption--corrected, $EF(E)$ spectra for the first
 ({\it left panel}) and second ({\it right panel}) observation of
 X1543--624 with superimposed the best fit model composed by a 
 \dbb\ ({\it dotted line}) plus \compbb\ ({\it dashed line}) plus a Gaussian
 ({\it dotted--dashed line}) at 6.4 keV
(see text). {\it Bottom panels}: residuals, in units of $\sigma$, to the 
best fit model photoelectrically--absorbed with non--standard values
 of O and Ne ({\sc vphabs}).}
\label{f:nuFnu_X1543-624}
\end{center}
\end{figure*}

Stimulated by the residuals in the 5--10 keV energy range during the first observation 
(see Fig.~\ref{folded_diskbbcompbb_x1543}), we investigated the presence of Fe K emission 
lines and/or K edges.  Assuming for the emission line a Gaussian profile
($I_l/ (\sqrt{2\pi} \sigma_l) \times exp[-(E-E_l)^2/ 2{\sigma_l}^2]$), with centroid energy $E_l$ in the range from 6.4 to 6.9 keV
the best fit was obtained with $E_l = 6.4$ keV, with a decrease in the \chiq/dof from
160/130 to 150/128 (significance level of $\sim$ 1\%) in the first observation and from
131/130 to 125/128 (significance level of $\sim$ 5\%) in the second observation.
The best fit parameter values of the line are reported in Table \ref{t:results}.
We note that these values are marginally consistent with those obtained by S94 and A2000.
The $E F(E)$ unabsorbed spectrum of the source with its  best fit
model, components and residuals of the data to the model is shown in 
Fig.~\ref{f:nuFnu_X1543-624} for both observations.

Adding an absorption edge in both observations did not improve 
the fit and the optical depth  is consistent with zero. 
As can be seen from Table~\ref{t:results}, the best fit values of $N_{\rm H}$ in the
two observations are marginally consistent with each other and with the Galactic value obtained
from the radio data along the source direction ($3 \times 10^{21}$~cm$^{-2}$,
Dickey \& Lockmann 1990 \nocite{Dickey90}). 

The source X--ray unabsorbed
 luminosity in various energy bands is shown in Table~\ref{t:results}.  
 While the 1--20 keV luminosity shows a very slight  decrease ($\sim$10\%) 
 from the first to the second observation, the partially extrapolated 20--200 keV   luminosity increases by a factor of about 2.

%
%Figure 4
%

%
% Figure 5
%

%
%Figure 6
%
\begin{figure*}
\begin{center}
\epsfig{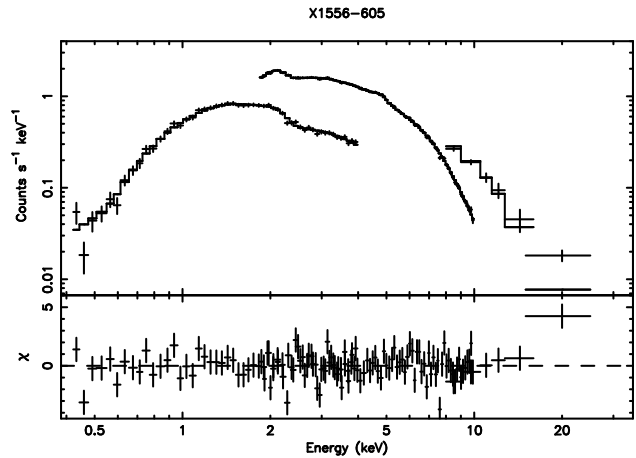}
\end{center} 
%\vspace{-6cm}
\caption[]{{\it Top panel}: average count rate  spectrum of the two 
observations of X1556--605 and best fit folded model consisting of 
a {\sc bb} plus a {\sc dbb} photoelectrically--absorbed (\wabs). 
{\it Bottom panel}: residuals to the 
model in units of $\sigma$. The excess above 15 keV can be recovered
 either replacing the \bb\ with a Comptonized \bb\ or replacing the \dbb\ with 
an \usc\ model (see text).} 
\label{f:sp_1556}
\end{figure*}

%
% Figure 7
%

%
% Figure 8
%
\begin{figure*}
\begin{center}
\epsfig{figure=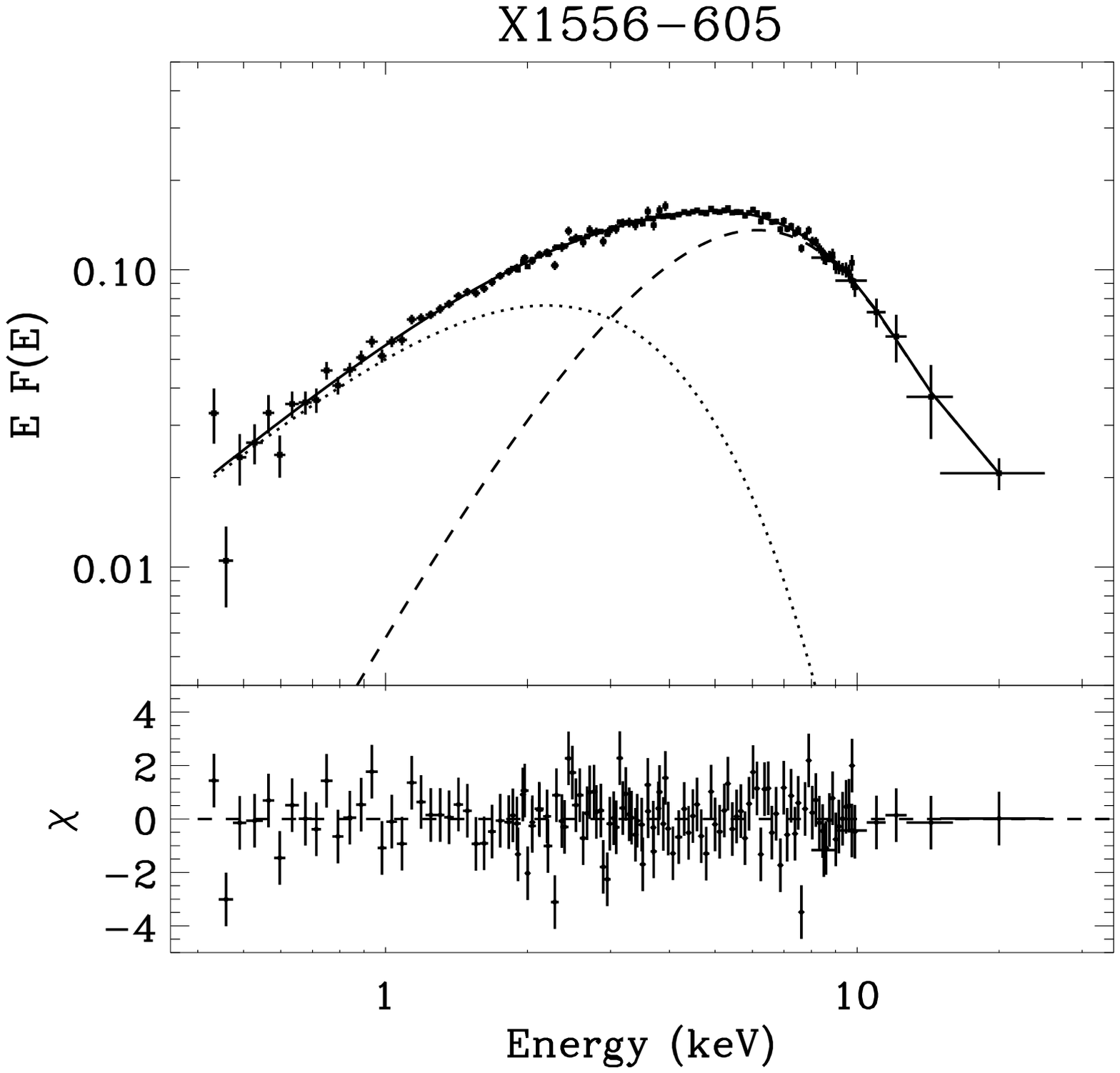,height=6cm,width=6.5cm}
\epsfig{figure=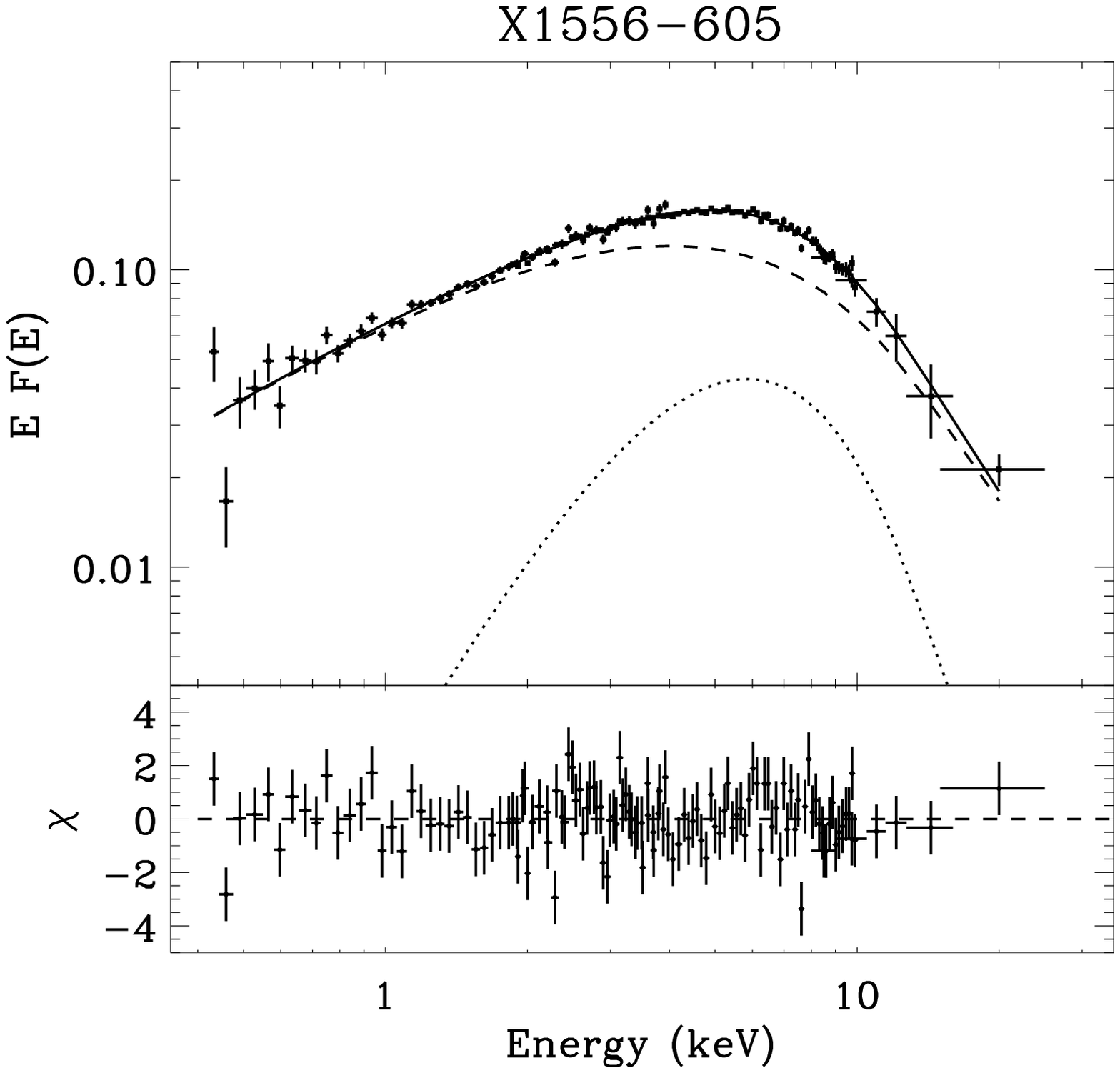,height=6cm,width=6.5cm} 
\vspace{0.5cm}
\caption[]{Same as in Fig. \ref{f:nuFnu_X1543-624} but for the time averaged
 spectrum of X1556--605. {\it Left panel}: \dbb\ ({\it dotted line}) plus \compbb\
 ({\it dashed line}). {\it Right  panel}: \bb\ ({\it dotted line}) plus \usc\
 ({\it dashed line}). {\it Bottom panels}: residuals, in units of $\sigma$,
to the best fit models, photoelectrically--absorbed (\wabs).}
\label{f:nuFnu_X1556-605}
\end{center}
\end{figure*}

\subsubsection{X1556--605}
\label{ss:1556}
 Given the lower statistical quality of the spectra of X1556--605 (see
count rate from the source in Table~\ref{t:log}),
the spectral analysis of the single observations could not be as detailed
as in the case of X1543--624. The count rate spectra of
the two observations, within their uncertainties, were consistent with
each other, so we performed the spectral analysis on the average spectrum.
A photoelectrically--absorbed \pl\ is unable to describe the data
($\chi^2_{\nu} > 10$). Models with a high--energy cut--off ({\sc usc} and {\sc compst} in order to 
compare our results with those obtained by M89) are still unsatisfactory 
(with {\sc usc} \chiq/dof = 160/126, while with \compst\ \chiq/dof=183/126), but the best fit 
parameter values are very similar to those obtained by M89. 
The photoelectrically--absorbed  two--component model \dbb\ plus \bb\ also leaves 
a significant excess above 15 keV (\chiq/dof =156/125, see Fig.~\ref{f:sp_1556}); instead
the photoelectrically--absorbed \bb\ plus \usc\ or \dbb\ plus \compbb\ give satisfactory
descriptions of the data (\chiq/dof = 141/124 and  134/123, respectively).
The unabsorbed $E F(E)$ model spectra and the residuals to the models are shown in 
Fig.~\ref{f:nuFnu_X1556-605}, while the best fit parameters 
are reported in Table~\ref{t:results}.
 
The derived $N_{\rm H}$ is in agreement
with the Galactic column density $N_{\rm H} = 0.3 \times10^{22}$ cm$^{-2}$
along the source direction derived from radio maps \cite{Dickey90} and with the
value ($0.27\times10^{22}$ cm$^{-2}$) derived from the extinction $E_{B-V}=0.55$ of 
the optical counterpart by M89, adopting the $E_{B-V}$ vs. $N_{\rm H}$ relation by 
Diplas \& Savage (1994)\nocite{Diplas94}.
We find no evidence of emission lines in the spectrum.
The luminosities in the same energy ranges adopted for X1543--624
 are reported in Table~\ref{t:results}.
 As can be seen, the source luminosity derived with the two models are
 similar in the 1--20 keV energy band.
 However, the partially extrapolated 20--200 keV luminosity is model dependent:
 it is higher in the case of  \bb\ plus \compbb\ than
in the case of \bb\ plus {\sc usc}.

\subsection{Timing properties}

No periodicity down to $\sim 10$~ms time scale can be inferred from the
power spectral density (PSD) estimate of the 2--10~keV light curves 
 (MECS data) of the two sources.
The PSD, evaluated in $\sim 10^{-3}$--10 Hz frequency range, do not show
statistically significant fractional variations of the source
fluxes: in the case of X1543--623 the  upper limit at 95\% confidence level
is $\sim$9\% in the first observation and 12\% in the second one,
in the case of X1556--605 the  upper limit at 95\% confidence level
is 14\% in the first observation and 16\% in the second one.

%
% Figure 9 
%
\begin{figure*}
\begin{center}
\epsfig{figure=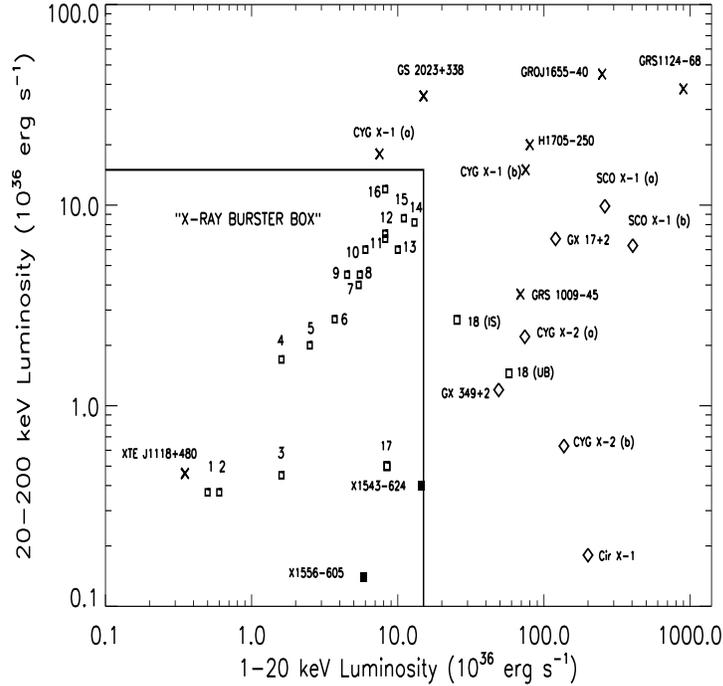,width=10cm,height=10cm}
\end{center} 
%\vspace{-6cm}
\caption[]{The hard (20--200 keV) X--ray luminosity versus the
soft (1--20 keV) luminosity for weakly magnetic compact X--ray binaries.
{\it Open squares}: X-ray bursters detected up to 100 keV and listed in
the analoguous figure by B2000; {\it filled circles}:
atoll sources in their 'island' or upper 'banana' states, when known;
{\it open diamonds}: Z sources;  {\it crosses}: black holes candidates. The
{\it `'X--ray burster box''} introduced by B2000
\nocite{Barret00} is also shown. BHCs are mostly
located on the right, but one of them (XTE J1118+480) is inside the X--ray
burster box. The position of our two sources ({\it filled squares}) within
the X--ray burster box is shown. The  points ({\it a}) and ({\it b})
for Sco X--1 denote
 the minimum and maximum fluxes observed by D'Amico et al. (2001) during 
their detection of a hard tail; the  points ({\it a}) and ({\it b}) for 
Cyg X--2  denote the cases in which the the hard X--ray tail was detected
and not detected, respectively (Di Salvo et al. 2002). The points
({\it a}) and ({\it b}) for Cyg~X--1 denote the low/hard state and the
high/soft state, respectively. The numbers from 1 to 16 denote the X--ray
bursters listed by B2000\nocite{Barret00}, while 17 and 18
denote two additional atoll sources included in the figure;
1: Aql X--1; 2: SLX1732--304; 3: 4U0614+09; 4: XB1323--619; 5: SAXJ1808.4--3658;
6: 4U1915--05; 7: SLX1735--269; 8: A1742--294; 9: 4U1608--52; 10: 
SAXJ1748.9--2021; 11: 1E1724--3045; 12: GS1826--238; 13: KS1731--260;  
14: GX354--0; 15: Cen X--4; 16: 4U1705--44; 17: 4U1728--34 \cite{Disalvo00b}; 
18: 4U1820-30 \cite{Bloser00}.}
\label{f:barret}
\end{figure*}

\section{Discussion}
\label{s:disc}
It is the first time that two unclassified LMXB sources,
X1543--624 and X1556--605, have been deeply investigated in a broad energy
band (0.4--200 keV) with two observations separated by 
about one month. The first source was detected up to 50 keV and the
other up to 25 keV. Both sources show many similar properties. We discuss now
the implications of these properties.

\subsection{Source nature and classification}
\label{s:disc1}
 The spectrum of  both X1543--624 and X1556--605
can be described by a two--component model consisting of a soft component
 ({\sc dbb}), likely coming from a cool accretion disk, plus a hard component 
 provided by the Comptonization of a {\sc bb}. The  \bb\ temperature $\ktbb$ of both 
sources is $\sim 1.5$ keV, while their \dbb\ temperature $\ktin$ ranges from $\sim$ 0.6 
to $\sim$ 0.9
keV. Both $\ktbb$ and other parameters of the model ($\kte$, $\rin \sqrt {cos(i)}$)
are in the range of values generally found
for LMXBs containing a low--magnetic field NS. Only the best fit value of $\tau$ ($\la$ 1), 
for both sources, is lower than the typical range of values (5--15) found for
this class of sources (see review by Barret 2001\nocite{Barret01}). 
However the best fit value of this parameter is model dependent (e.g., we find
$\tau > 10$ for a \bb\ plus \comptt\ model).
All that strongly points to a NS nature of both X1543--624 and X1556--605. The projected
inner disk radius obtained for X1543--624 is consistent with this conclusion, 
while a very small value ($\sim 4$ km) is found for X1556--605. However, small values 
of the projected inner disk radius have been found in other weakly 
magnetized accreting NS (e.g., 2.8 km for GX3+1, Oosterbroek et al. 
2001\nocite{Oosterbroek01}). Actually the {\it effective} inner disk radius
 \cite{ST95,Merloni00} is given by  $\reff \sim \rin \times f_{\rm col}^2$, 
where $f_{\rm col}$ is the spectral hardening factor. As discussed by
Shimura \& Takahara (1995), for luminosities
down to $\sim$ 0.1 times the Eddington luminosity, the conventional 
value $f=1.7$ can be adopted. This condition can be applied to X1543--624 ($L_{\rm 0.1-200~keV} \sim 0.1 \times L_{Edd}$, see below), finding $\reff \sim 60$ km for a disk inclination angle $i=60^{\circ}$ (the source does not show dips or eclipses) and,
marginally, for X1556--605 ($L_{\rm 0.1-200~keV} \sim 0.05 \times L_{Edd}$ for
a distance of 10 kpc, see below), finding $R_{\rm eff} \sim 16$ km for the
same disk inclination angle. 
 
A common property of our sources is their flux and spectrum stability. 
From the ASM data archive\footnote{ASM archive is available by
 internet at {\it http://xte.mit.edu/XTE/asmlc/ASM.html}} one can see that 
both sources show intensity variations not greater than a factor of 2 on 
time scales of months. As shown in Table~\ref{t:results}, the parameter values of the best fit spectrum of X1543--624 also do not change from
the first to the second observation, except the plasma electron temperature $\kte$ 
which increases from $\sim 7$ to $\sim 25$ keV. The spectra 
obtained in the two observations of X1556--605 
are consistent with each other, as discussed in Section~\ref{ss:1556}. 
On longer time scales the source spectra appear
stable: when we adopted the same models used to describe the source spectra
obtained in the previous observations
 (a \bb\ plus \dbb\ for X1543--624, A2000; a simple \usc\ or \compst\
 for X1556--605, M89) the fits, even if unsatisfactory, provided similar parameter values.

The spectral stability of both sources is also apparent from their CD
 (see Fig.~\ref{f:cd}), which is unlike those of 
 traditional Z sources, which show spectral variations on time scales
 of hours to days. Both the soft and hard colours change
 by about  20\%, an extent  consistent with that exhibited
by atoll sources in their banana branch. The CD shape is similar to that of  atoll
sources with low flux variations ($F_{max}/F_{min} < 10$) \cite{Muno02}.
A similar CD pattern is shown, e.g., by the X--ray burster
GS1826--238, which also exhibits
intensity variations within a factor of 2 \cite{Muno02}. The only difference
we find with GS1826--238 is in the centroid value of the hard colour, which
in the latter source  is  significantly higher  ($\sim 1.5$), as found
with \sax\ \cite{Delsordo02}.

The luminosity of  X1543--624 and X1556--605 supports an atoll
classification. In the case of  X1543--624, the average bolometric
luminosity (0.1--200 keV) is $L_X \sim 2 \times 10^{37}
(d/10~{\rm kpc})^2$ erg s$^{-1}$, which corresponds to $\sim 0.1
\times L_{{\rm Edd}}$ (the Eddington luminosity is $L_{Edd} =
1.48 \times 10^{38} M/M_{\odot}$ erg s$^{-1}$  if an H mass fraction
$f_H = 0.7$ and a NS mass of $1.4 M_{\odot}$ is assumed), and, in the
case of X1556--605 it is $L_X \sim 7 \times 10^{36}
(d/10~{\rm kpc})^2$~erg s$^{-1}$ corresponding to $\sim 0.05 \times L_{Edd}$.
Both luminosities are typical of X--ray bursters (see e.g., Barret et al. 
2000\nocite{Barret00}, hereafter B2000).

We have also checked the atoll assumption using the luminosity comparison
criterion first introduced by Barret et al. (1996)\nocite{Barret96}.
In Fig.~\ref{f:barret} we show the hard X--ray (20--200~keV) luminosity  
versus the soft X--ray (1--20~keV) luminosity not only for the X--ray
bursters and BHCs candidates listed by B2000\nocite{Barret00},
but also for Z sources during their exhibition of hard tails  (see Section~\ref{s:intro}), 
for a new transient BHC (XTE J1118+480, Frontera et al. 2001\nocite{Frontera01}), and
for two additional atoll sources, 4U1728--34 and 4U1820--30, both detected up to 100 keV,
the first with \sax\ \cite{Disalvo00b}, the second with RXTE \cite{Bloser00} 
in both the 'island' and upper 'banana' states.
As can be seen, X1543--624 and X1556--605 are both in the {\it X--ray
burster box} introduced by Barret et al. (1996), but this fact does not 
help to constrain the source class. Indeed, from one side,  XTE J1118+480 is 
inside this box, while, on the other side,  an atoll, 4U1820--30, 
is outside the box in both the island and banana states, moreover in positions which are 
indistinguishible from Z sources with hard tails and from some of BHCs.
Note that also the Z sources Cyg X-2 \cite{Kuulkers95}
 and GX17+2 (Sztajno et al. 1986, Kuulkers et al. 1997) are X--ray bursters,
 and are outside the X--ray burster box.

If X1543--624 and X1556--605 are atolls, the absence of broad--band noise,
as found from the PSD estimate, would be in favour of a banana state for these
sources.

 An open question is the absence of X--ray bursts from the two sources,
which would definitively confirm the NS nature of the compact object.
However the fact that bursts have not been observed does not imply that they
do not occur; there is still not an extensive coverage of both X1543--624 and 
X1556--605 (either with \sax or with other satellites), so bursts can
be easily missed, especially if both sources have low burst
recurrence. 
In fact there are sources which, even though they have been known for many years, only
recently have shown X--ray bursts (SLX 1737--282, in 't Zand 
et al. 2002\nocite{Zand02};  2S 1711--339, Cornelisse et al. 2002\nocite{Cornelisse02}).
In the case of X1543--624, an unusual composition of the companion might
lead to atypical burst properties, such as very long recurrence times ($> 0.5$ yr for
thermonuclear flashes in pure C layers)  and corresponding 
large fluences \cite{Joss80}. Three of the four sources analyzed by J2001 
(4U 0614+091, Swank et al. 1978\nocite{Swank78}, Brandt et al. 1992\nocite{Brandt92}; 
2S 0918-549, Jonker et al. 2001\nocite{Jonker01}; 4U 1850--087, Hoffman et al. 1980\nocite{Hoffman80}) have shown burst
properties which induced the authors to exclude the possibility of the
donor being a C--O dwarf. However we suggest that the conclusion of J2001
cannot be extended to X1543--624, because this source, unlike  
  the three sources mentioned above, still has not shown X--ray bursts
whose properties could exclude a C--O dwarf nature of the companion.
In the case of X1556--605, the burst absence (or recurrence 
on very long time scales) could be due to an underabundance of CNO,
 which was observed in this source by M89.
 CNO is an important parameter which determines the rate at which hydrogen burns
into helium \cite{Hoyle65} and therefore the recurrence time 
of X--ray bursts \cite{Fujimoto87}. Moreover in X1556--605 the accretion rate 
is low, as inferred by its X--ray luminosity. Wallace et al. (1982)
 have shown that low metallicity in conjunction with a low accretion rate
 favours the long interval between bursts and very energetic thermonuclear
 runaways, and this could be the case of X1556--605 (M89).

\subsection{Accretion geometry}
\label{s:disc2}
 Including the corrections for the spectral hardening factor we found that
the inner disk radius of X1543--624 should be at $\sim 5$ NS radii.
A likely explanation of this result, taking into account the low magnetic field
of the source class to which X1543--624 belongs, is that the inner accretion 
disk is replaced by an optically thin hot accretion flow, following  
the scenario suggested by B2000 on the basis of the spectral properties of a
sample of bursters studied with {\it RXTE}.
Optically thin hot accretion flows (the {\it Advection Dominated Accretion
Flows}, ADAF) have been demonstrated \cite{Narayan95} to be
stable and preferrably occur at relatively low accretion rates 
(such as in the case of  X1543--624) and this scenario strengthens the atoll 
hypothesis for this source. One controversial point is that the spectrum 
hardens 
(as testified by the significant increase of $\kte$, see 
Table~\ref{t:results}) despite the fact that the \dbb\ luminosity increases, while
the 0.1--200 keV luminosity remains constant.
Another anomalous behaviour has also been noticed by Schultz (2002), as discussed in
Section~\ref{X1543--624:l}: a spectral hardening  as the source luminosity 
increases.
The reason for this different behaviour is unclear and could
be connected to a quite complex geometry of the accretion flow.

In the case of X1556--605, the inner disk radius, once corrected for the 
spectral hardening factor, is consistent with being close to the 
NS surface, so that
for this source the ADAF scenario appears less suitable.
On the other hand, on the basis of the above considerations,
it is  unclear why the ADAF scenario should occur in X1543--624 and not 
in X1556--605, which is  less luminous and thus is expected to have
a lower accretion rate. 
However, given that the X--ray spectrum of X1556--605 can be
also described by a \bb\ plus \usc\ (Western model), another accretion 
scenario can be considered for this source, in which the soft component 
(\bb\ with $\ktbb \sim 1.5$~keV) arises from the NS surface (or the boundary layer) 
and the hard component is
due to Comptonization of the radiation from the inner disk.
In this scenario the small \bb\ radius derived ($\sim 1$ km, see Table~\ref{t:results}) 
could be ascribed to non--isotropic emission from the NS, as in the case of 
emission coming from an equatorial belt \cite{IS99}, 
or to the fact that the plasma responsible for the Comptonization
of the radiation from the inner disk also intercepts part of the flux coming from the
 NS surface, thus reducing the \bb\ flux and hence its radius. 

The evidence of an Fe K emission line from X1543--624, with the energy 
centroid of the line at 6.4 keV, suggests
fluorescence emission from a neutral medium, such as from a disk not extending
down to the NS surface, in agreement with the above considerations. However
the non--detection of an Fe K line from X1556--605 remains unclear if the disk
extends down to the NS surface.
 
\section{Conclusions}
\label{s:concl}
From the above discussion it emerges that X1543--624 and X1556--605
show properties of atolls in the banana state, even if the shape of the
CD is not strictly reminiscent of a banana. However it
is similar to the CD of atoll sources with small
intensity variations (e.g., GS1826$-$238). The high
$L_{\rm 1-20\,keV}/L_{\rm 20-200\,keV}$  ratio is also found
in atoll sources (see Fig~\ref{f:barret}).
The absence of type I X--ray bursts may be explained by their long
recurrence times and by the short observation times of the two sources.
The accretion geometry of X1543--624 is consistent with that of sources 
 inside the X--ray burster box, for which
an ADAF scenario appears suitable to describe their spectral behaviour
(B2000). 
X1543--624 shows, in addition to an Fe K emission line, a stable
emission feature at $\sim$0.7 keV, which is likely
due to the K edges of local O and Ne with non--solar abundances, as
found for other sources (J2001). Even in this  case X--ray
bursts, although with atypical properties and frequencies, should be
expected, but they have not been observed until now.
For X1556--605 both the scenarios proposed by the Western  and 
Eastern models are possible, and on the basis of our data we cannot
decide which of them is preferrable.
A broad--band, long and more sensitive monitoring   could be of
key importance to better understand these  LMXBs and to 
better constrain the mass accretion scenario.

\acknowledgements
Many thanks to the anonymous referee for the very useful suggestions
and comments on the first version of our paper.
This research is supported by the Italian Space Agency (ASI) and
Ministry of University and Scientific Research of Italy (COFIN funds).
\sax\ is a joint Italian and Dutch program.


\begin{thebibliography}{}

\bibitem[Anders \& Grevesse 1989]{Anders89}
Anders, E. \& Grevesse, N. 1989, Geochim. Cosmochim. Acta 53, 197
%
\bibitem[Apparao et al.\  1978]{Apparao78}
Apparao, K.M.V., Bradt, H.V., Dower, R.G., et al. 1978, Nat 271,
	225
%
\bibitem[Arnaud 1996]{Arnaud96}
Arnaud, K.A. 1996, XSPEC: the first ten years. In: Jacoby G.H.,
        Barnes J. (eds.) ``Proceedings of the V ADASS Symposium",
        ASP Conf. Ser. 101, 17
%
\bibitem[Asai et al. 2000]{Asai00}
Asai, K., Dotani, T., Nagase, F. \& Mitsuda, K. 2000, ApJS 131, 571 (A2000)
%
\bibitem[Asai et al.\ 1994]{Asai94}
Asai, K., Dotani, T., Mitsuda, K. et al. 1994, PASJ 46, 479
%
\bibitem[Barret et al.\ 1996]{Barret96}
Barret, D., McClintock, J. E. \& Grindlay, J.E., 1996, ApJ 473, 963
% 
\bibitem[Barret et al.\ 2000]{Barret00}
Barret, D., Olive, J. F., Boirin, L. et al. 2000, ApJ 533, 329 (B2000)
%
\bibitem[Barret\ 2001]{Barret01}
Barret, D. 2001, AdSpR 28, 307
%
\bibitem[Bloser et al.\ 2000]{Bloser00}
Bloser, P. F., Grindlay, J. E., Kaaret, P. et al. 2000, ApJ 542, 1000
% 
\bibitem[Boella et al.\ 1997a]{Boella97a}
Boella, G., Butler, R.C., Perola, G.C. et al. 1997a, A\&AS 122,
	299
%
\bibitem[Boella et al.\ 1997b]{Boella97b}
Boella, G., Chiappetti., L., Conti., G. et al. 1997b, A\&AS 122,
	327
%
\bibitem[Bradt \& McClintock 1983]{Bradt83}
Bradt, H.V. \& McClintock, J.E. 1983, ARA\&A 21, 13
%
\bibitem[Brandt et al. 1992]{Brandt92}
Brandt, S., Castro--Tiraldo, A. J., Lund, N. et al. 1992, A\&A 262, L15
%
\bibitem[Charles et al.\ 1979]{Charles79}
Charles, P.A., Thorstensen, J.R., Bowyer, S., et al. 1979, BAAS
	11, 720
%
\bibitem[Chiappetti \& Dal Fiume 1997]{Chiappetti97}
        Chiappetti, L. \& Dal Fiume, D. 1997, in: ``Proceedings of the Fifth
        International Workshop on Data
        Analysis in Astronomy" System, Eds. di Ges\`u V., Duff M.J.B.,
        Heck A., Maccarone M.C., Scarsi L., Zimmermann H.U. (World
        Scientific Press), p. 101
%
\bibitem[Christian \& Swank 1997]{Christian97}
Christian, D.J. \& Swank, J.H. 1997, ApJS 109, 177
%
\bibitem[Cornelisse et al. 2002]{Cornelisse02}
Cornelisse, R., Verbunt, F, in 't Zand, J.J.M. et al. 2002, A\&A 392, 885
%

\bibitem[D'Amico et al.\ 2001]{Damico01}
D'Amico, F., Heindl, W. A., Rothschild, R. E. et al. 2001, ApJ 547, L147
%
\bibitem[Del Sordo et al.\ 2002]{Delsordo02}
Del Sordo, S. et al., in preparation (2002)
%
\bibitem[Dickey \& Lockmann 1990]{Dickey90}
Dickey, J.M. \& Lockmann, F.J. 1990, ARA\&A 28, 215
% 
\bibitem[Diplas \& Savage 1994]{Diplas94}
Diplas, A. \& Savage, B. D., 1994, ApJ 427, 274
%
\bibitem[Di Salvo et al.\ 2000a]{Disalvo00a}
Di Salvo, T., Stella, L., Robba, N. R. et al. 2000a, ApJ 544, L119
%
\bibitem[Di Salvo et al.\ 2000b]{Disalvo00b}
Di Salvo, T., Iaria, R., Burderi, L. \& Robba, N. R. 2000b, ApJ 542, 1034
%
\bibitem[Di Salvo et al.\ 2001]{Disalvo01}
Di Salvo, T., Robba, N. R., Iaria, R. et al. 2001, ApJ 554, 49
%
\bibitem[Di Salvo et al.\ 2002]{Disalvo02}
Di Salvo, T., Farinelli, R., Burderi, L. et al. 2002, A\&A 386, 535
%

 \bibitem[Fiore et al.\ 1999]{Fiore99}
        Fiore, F., Guainazzi, M. \& Grandi, P., 1999, Technical Report 1.2,
        {\it BeppoSAX} scientific data center, available online at {\tt
        ftp://www.sdc.asi.it/pub/sax/doc/\\
	software\_docs/saxabc\_v1.2.ps}
%
\bibitem[Frontera et al.\ 1997]{Frontera97}
Frontera, F., Costa, E., Dal Fiume, D. et al. 1997, A\&AS 122,
	357
%
\bibitem[Frontera et al.\ 1998]{Frontera98}
Frontera, F., Dal Fiume, D., Malaguti, G. et al. 1998, 
in: ``The Active X--ray Sky'', eds. L. Scarsi, H. Bradt, 
P. Giommi \& F. Fiore, Nucl. Phys. B, 69, 286
%
\bibitem[Frontera et al.\ 2001]{Frontera01}
Frontera, F., Zdziarski, A.A., Amati, L. et al. 2001, ApJ 561, 1006
%
\bibitem[Fujimoto et al.\ 1987]{Fujimoto87}
 Fujimoto, M. Y., Sztajno, M., Lewin, W. H. G. et al. 1987, ApJ 319, 902
%
\bibitem[Gierli\'nski \& Done 2002]{Gierlinski02}
Gierli\'nski, M. \& Done, C. 2002, MNRAS 331, L47 
%
\bibitem[Gottwald et al.\ 1995]{Gottwald95}
Gottwald, M., Parmar, A.N., Reynolds, A.P., et al. 1995, A\&AS 109, 9
%
\bibitem[Harmon et al. 1996]{Harmon96}
Harmon, B. A., Wilson, C. A., Tavani, M. et al. 1996, A\&AS 120, 197
%
\bibitem[Hasinger \& van der Klis 1989]{Hasinger89}
 Hasinger, G. \& van der Klis, M. 1989, A\&A 225, 79
%
\bibitem[Hoffman et al. 1980]{Hoffman80}
Hoffman, J.A., Cominsky, L. \& Lewin, W. H. G. 1980, ApJ 240, L27
%
\bibitem[Hoyle \& Fowler \ 1965]{Hoyle65}
Hoyle, F. \& Fowler, W. A. 1965, in {\it Quasi--stellar Sources and Gravitational
 Collapse}, eds. I. Robinson, A. Schild, E. L. Shucking, University of Chicago Press,
 Chicago, p. 17
%
\bibitem[Iaria et al.\ 2001]{Iaria01}
         Iaria, R., Burderi, L., Di Salvo, T. et al. 2001, ApJ 547, 412
%
\bibitem[in 't Zand et al.\ 2002]{Zand02}
 in 't Zand, J.J.M, Verbunt, F., Kuulkers, E. et al. 2002, A\&A L43, 389
%
\bibitem[Inogamov \& Sunyanev 1999]{IS99}
Inogamov, N. A. \& Sunyaev, R. A. 1999, AstrL 25, 269

\bibitem[Jonker et al. 2001]{Jonker01}
Jonker, P. G., Peter, G., van der Klis, M. et al. 2001, ApJ 553, 335

\bibitem[Joss \& Li 1980]{Joss80}
Joss, P.C. \& Li, F.K. 1980, ApJ 238, 287
%
\bibitem[Juett et al.\ 2001]{Juett01}
Juett, A.M., Psaltis, D. \& Chakrabarty, D. 2001, ApJ 560, 59 (J2001)
%
\bibitem[Kuulkers et al.\ 1995]{Kuulkers95}
Kuulkers, E., van der Klis, M. \& van Paradjis, J. 1995, ApJ 450, 748
%
\bibitem[Kuulkers et al.\ 1997]{Kuulkers97}
Kuulkers, E., van der Klis, M., Oosterbroek, T. et al. 1997, MNRAS 287, 495
%
\bibitem[Lammers 1997]{Lammers97}
Lammers, U. 1997, The SAX/LECS Data Analysis System User
	Manual, SAX/LEDA/0010
%
\bibitem[Manzo et al. 1997]{Manzo97}
Manzo, G., Giarrusso, S., Santangelo, A. et al. 1997, A\&AS 122,
	341
 %
\bibitem[McClintock et al.\ 1978]{McClintock78}
McClintock, J.E., Canizares, C., Hiltner, W.A. \& Petro, L. 1978,
	IAU Circ. 3251
%
\bibitem[Merloni et al. \ 2000]{Merloni00}
Merloni, A., Fabian, A. C. \& Ross, R. R. 2000, MNRAS 313, 193
%

\bibitem[Mitsuda et al.\ 1984]{Mitsuda84}
Mitsuda, K., Inoue, H., Koyama, K. et al. 1984, PASJ 36 741
%
\bibitem[Mitsuda et al.\ 1989]{Mitsuda89}
Mitsuda, K., Inoue, H., Nakamura, N. \& Tanaka, Y.  1989, PASJ 41, 97
%
\bibitem[Morrison \& McCammon 1983]{Morrison83}
Morrison, R. \& McCammon, D. 1983, ApJ 270, 119
%
\bibitem[Motch et al.\ 1989]{Motch89}
Motch, C., Pakull, M.W., Mouchet, M. \& Beuermann, K. 1989, A\&A
        219, 158 (M89)
%
\bibitem[Muno et al. 2002]{Muno02}
Muno, M.P., Remillard, R.A. \& Chakrabarty, D. 2002, ApJ 568, L35
%
\bibitem[Narayan \& Yi 1995]{Narayan95}
 Narayan, R. \& Yi, I. 1995, ApJ 452, 7101 
%
\bibitem[Nishimura et al. 1986]{Nishimura86}
Nishimura, J., Mitsuda, K. \& Itoh, M. 1986, PASJ 38, 819
%
\bibitem[Oosterbroek et al.\ 2001]{Oosterbroek01}
Oosterbroek, T., Barret, D., Guainazzi, M. \& Ford, E.C. 2001, A\&A 366, 138
%
\bibitem[Parmar et al.\ 1997]{Parmar97}
Parmar, A., Martin, D.D.E, Bavdaz, M. et al. 1997, A\&AS 122, 309
%
\bibitem[Piraino et al. 1999]{Piraino99}
Piraino, S., Santangelo, A., Kaaret., P. et al. 1999, A\&A 349, L77
%
\bibitem[Schultz 2002]{S02}
Schultz, J. 2002, astro--ph/0210250
%
\bibitem[Shimura \& Takahara  \ 1995]{ST95}
Shimura, T. \& Takahara, F. 1995, ApJ 445, 780
%
\bibitem[Singh et al.\ 1994]{Singh94}
Singh, K.P., Apparao, K.M.V. \& Kraft, R.P. 1994, ApJ 753, 761 (S94)
%
\bibitem[Smale 1991]{Smale91}
        Smale, A.P. 1991, PASP 103, 636
%
\bibitem[Smith et al.\ 1990]{Smith90}
Smith, H.A., Beall, J.H. \& Swain, M.R. 1990, AJ 99, 273
%
\bibitem[Sztajno et al. 1986]{Sztajno86}
Sztajno, M., van Paradijs, J., Lewin, W. H. G. et al. 1986, MNRAS 222, 499
%
\bibitem[Sunyaev \& Titarchuk 1980]{Sunyaev80}
         Sunyaev, R.A. \&  Titarchuk, L. 1980, A\&A 86, 121
%
\bibitem[Swank et al. \ 1978]{Swank78}
Swank, J .H., Becker, R. H., Boldt, E. A. et al. 1978, MNRAS 182, 349
% 
\bibitem[Titarchuk 1994]{Titarchuk94}
Titarchuk, L.G. 1994, ApJ 434, 570
%
\bibitem[Van der Klis et al.\ 1990]{Vanderklis90}
van der Klis, M., Hasinger, G., Damen, E. et al. 1990, ApJ 360, L19
%
\bibitem[van der Klis 1995]{Vanderklis95}
        van der Klis, M. 1995, in: ``X--ray Binaries'', eds. Lewin W.H.G.,
        van Paradijs J. \& van den Heuvel E.P.J. (Cambridge Univ. Press)
        p. 252
%
\bibitem[van Paradijs \& McClintock 1995]{Jvp95}
       van Paradijs, J. \& McClintock, J.E. 1995, in: ``X--ray Binaries'',
       eds. Lewin W.H.G., van Paradijs J. \& van den Heuvel E.P.J.
       (Cambridge Univ. Press) p. 536
%
\bibitem[Wallace et al. \ 1982]{Wallace82}
Wallace, R. K., Woosley, S. E. \&  Weaver, T. A. 1982, ApJ 258, 696
%
\bibitem[Warwick et al.\ 1981]{Warwick81}
        Warwick, R.S., Marshall, N., Fraser, G.W. et al. 1981, MNRAS
	197, 865
%
\bibitem[Wendker 1995]{Wendker95}
         Wendker, H.J. 1995, A\&AS 109, 177

%
\bibitem[White et al. 1986]{White86}
White, N. E., Peacock, A., Hasinger, G. et al. 1986, MNRAS 218, 129 


\bibitem[White et al 1988]{White88}
White, N.E., Stella, L. \& Parmar A.N. 1988, ApJ 324, 363

%
\bibitem[White et al. 1997]{White97}
White, N.E., Kallman, T.R., \& Angelini, L. 1997, in: "X--ray imaging
and Spectroscopy of Cosmic Hot Plasmas", eds. F. Makino and K. Mitsuda
(Tokio: Universal Academy Press), p. 411
%
\end{thebibliography}
\end{document}